\input harvmac
\catcode`\@11\relax
\newif\ify@autoscale \y@autoscaletrue \def\Yautoscale#1{\ifnum #1=0
  \y@autoscalefalse\else\y@autoscaletrue\fi}
\newdimen\y@b@xdim
\newdimen\y@boxdim \y@boxdim=13pt
\def\Yboxdim#1{\y@autoscalefalse\y@boxdim=#1}
\newdimen\y@linethick    \y@linethick=.3pt
\def\Ylinethick#1{\y@linethick=#1}
\newskip\y@interspace \y@interspace=0ex plus 0.3ex
\def\Yinterspace#1{\y@interspace=#1}
\newif\ify@vcenter   \y@vcenterfalse
\def\Yvcentermath#1{\ifnum #1=0 \y@vcenterfalse\else\y@vcentertrue\fi}
\newif\ify@stdtext   \y@stdtextfalse
\def\Ystdtext#1{\ifnum #1=0 \y@stdtextfalse\else\y@stdtexttrue\fi}
\newif\ify@enable@skew   \y@enable@skewfalse
\expandafter\ifx\csname enableskew\endcsname\relax
 \y@enable@skewfalse \else \y@enable@skewtrue\fi
\def\y@vr{\vrule height0.8\y@b@xdim width\y@linethick depth 0.2\y@b@xdim}
\def\y@emptybox{\y@vr\hbox to \y@b@xdim{\hfil}}
\ify@enable@skew
 \def\y@abcbox#1{\if :#1\else
   \y@vr\hbox to \y@b@xdim{\hfil#1\hfil}\fi}
 \def\y@mathabcbox#1{\if :#1\else
   \y@vr\hbox to \y@b@xdim{\hfil$#1$\hfil}\fi}
\else
 \def\y@abcbox#1{\y@vr\hbox to \y@b@xdim{\hfil#1\hfil}}
 \def\y@mathabcbox#1{\y@vr\hbox to \y@b@xdim{\hfil$#1$\hfil}}
\fi
\def\y@setdim{%
  \ify@autoscale%
   \ifvoid1\else\typeout{Package youngtab: box1 not free! Expect an
     error!}\fi%
   \setbox1=\hbox{A}\y@b@xdim=1.6\ht1 \setbox1=\hbox{}\box1%
  \else\y@b@xdim=\y@boxdim \advance\y@b@xdim by -2\y@linethick
  \fi}
\newcount\y@counter
\newif\ify@islastarg
\def\y@lastargtest#1,#2 {\if\space #2 \y@islastargtrue
  \else\y@islastargfalse\fi}
\def\y@emptyboxes#1{\y@counter=#1\loop\ifnum\y@counter>0
  \advance\y@counter by -1 \y@emptybox\repeat}
\def\y@nelineemptyboxes#1{%
  \vbox{%
    \hrule height\y@linethick%
    \hbox{\y@emptyboxes{#1}\y@vr}
    \hrule height\y@linethick}\vskip-\y@linethick}
\def\yng(#1){%
  \y@setdim%
  \hskip\y@interspace%
  \ifmmode\ify@vcenter\vcenter\fi\fi{%
  \y@lastargtest#1,
  \vbox{\offinterlineskip
    \ify@islastarg
     \y@nelineemptyboxes{#1}
    \else
     \y@ungempty(#1)
    \fi}}\hskip\y@interspace}
\def\y@ungempty(#1,#2){%
  \y@nelineemptyboxes{#1}
  \y@lastargtest#2,
  \ify@islastarg
   \y@nelineemptyboxes{#2}
  \else
   \y@ungempty(#2)
  \fi}
\def\y@nelettertest#1#2. {\if\space #2 \y@islastargtrue
  \else\y@islastargfalse\fi}
\def\y@abcboxes#1#2.{%
  \ify@stdtext\y@abcbox#1\else\y@mathabcbox#1\fi%
  \y@nelettertest #2.
  \ify@islastarg\unskip%
   \ify@stdtext\y@abcbox{#2}\else\y@mathabcbox{#2}\fi%
  \else\y@abcboxes#2.\fi}
 \newdimen\y@full@b@xdim
 \newcount\y@m@veright@cnt
\ify@enable@skew
 \def\y@get@m@veright@cnt#1#2.{%
   \if :#1 \advance\y@m@veright@cnt by 1\y@get@m@veright@cnt#2.\fi}
 \let\y@setdim@=\y@setdim
 \def\y@setdim{%
   \y@setdim@ \y@full@b@xdim=\y@b@xdim
   \advance\y@full@b@xdim by 1\y@linethick}
 \def\y@m@veright@ifskew#1{
   \y@m@veright@cnt=0 \y@get@m@veright@cnt#1.
   \moveright \y@m@veright@cnt\y@full@b@xdim}
\else
 \def\y@m@veright@ifskew#1{}
\fi
\def\y@nelineabcboxes#1{%
  \y@nelettertest #1.
  \ify@islastarg
   \y@m@veright@ifskew{#1}
    \vbox{
      \hrule height\y@linethick%
      \hbox{\ify@stdtext\y@abcbox#1\else\y@mathabcbox#1\fi\y@vr}
      \hrule height\y@linethick}\vskip-\y@linethick
  \else
   \y@m@veright@ifskew{#1}
    \vbox{
      \hrule height\y@linethick%
      \hbox{\y@abcboxes #1.\y@vr}%
      \hrule height\y@linethick}\vskip-\y@linethick
  \fi}
\def\young(#1){%
  \y@setdim%
  \hskip\y@interspace%
  \y@lastargtest#1,
  \ifmmode\ify@vcenter\vcenter\fi\fi{%
  \vbox{\offinterlineskip
    \ify@islastarg\y@nelineabcboxes{#1}%
    \else\y@ungabc(#1)%
    \fi}}\hskip\y@interspace}
\def\y@ungabc(#1,#2){%
  \y@nelineabcboxes{#1}%
  \y@lastargtest#2,
  \ify@islastarg\y@nelineabcboxes{#2}%
  \else\y@ungabc(#2)%
  \fi}
\catcode`\@12\relax
 

 \def\quad{{\ \ }}

\def\none{n_1}
\def\ntwo{n_2}
\def\nthree{n_3}
\def\nen{n_{{}_{N}}}

\let\includefigures=\iftrue
\newfam\black
\includefigures
\input epsf
\def\figin{\epsfcheck\figin}\def\figins{\epsfcheck\figins}
\def\epsfcheck{\ifx\epsfbox\UnDeFiNeD
\message{(NO epsf.tex, FIGURES WILL BE IGNORED)}
\gdef\figin##1{\vskip2in}\gdef\figins##1{\hskip.5in}
\else\message{(FIGURES WILL BE INCLUDED)}%
\gdef\figin##1{##1}\gdef\figins##1{##1}\fi}
\def\DefWarn#1{}
\def\figinsert{\goodbreak\midinsert}
\def\ifig#1#2#3{\DefWarn#1\xdef#1{fig.~\the\figno}
\writedef{#1\leftbracket fig.\noexpand~\the\figno}%
\figinsert\figin{\centerline{#3}}\medskip\centerline{\vbox{\baselineskip12pt
\advance\hsize by -1truein\noindent\footnotefont{\bf Fig.~\the\figno:}
#2}}
\bigskip\endinsert\global\advance\figno by1}
\else
\def\ifig#1#2#3{\xdef#1{fig.~\the\figno}
\writedef{#1\leftbracket fig.\noexpand~\the\figno}%
#2}}
\global\advance\figno by1}
\fi


\def\sym{  \> {\vcenter  {\vbox
                  {\hrule height.6pt
                   \hbox {\vrule width.6pt  height5pt
                          \kern5pt
                          \vrule width.6pt  height5pt
                          \kern5pt
                          \vrule width.6pt height5pt}
                   \hrule height.6pt}
                             }
                  } \>
               }
\def\fund{  \> {\vcenter  {\vbox
                  {\hrule height.6pt
                   \hbox {\vrule width.6pt  height5pt
                          \kern5pt
                          \vrule width.6pt  height5pt }
                   \hrule height.6pt}
                             }
                       } \>
               }
\def\anti{ \>  {\vcenter  {\vbox
                  {\hrule height.6pt
                   \hbox {\vrule width.6pt  height5pt
                          \kern5pt
                          \vrule width.6pt  height5pt }
                   \hrule height.6pt
                   \hbox {\vrule width.6pt  height5pt
                          \kern5pt
                          \vrule width.6pt  height5pt }
                   \hrule height.6pt}
                             }
                  } \>
               }

\lref\YamaguchiTE{
  S.~Yamaguchi,\hskip-3pt
  ``Bubbling geometries for half BPS Wilson lines,''
  arXiv:hep-th/0601089.
}

\Title{\vbox{\baselineskip12pt\hbox{hep-th/0604007}\hbox{KUL-TF-06/11}}}
{\vbox{\centerline{
Holographic Wilson Loops}}}

\centerline{Jaume Gomis\foot{jgomis@perimeterinstitute.ca} and Filippo Passerini\foot{Filippo.Passerini@fys.kuleuven.be}}
\medskip\medskip

\bigskip\centerline{\it Perimeter Institute for Theoretical Physics}
\centerline{\it Waterloo, Ontario N2L 2Y5, Canada$^{1,2}$}
\centerline{{\it and}}
\centerline{\it Instituut voor Theoretische Fysica, Katholieke Universiteit Leuven }
\centerline{\it Celestijnenlaan 200D B-3001 Leuven, Belgium$^{2}$}
\vskip .3in

\centerline{Abstract}

We show that all half-BPS Wilson loop operators in ${\cal N}=4$ SYM -- which are labeled by   Young tableaus --  have a gravitational dual description in terms of  $D5$-branes or alternatively in terms of $D3$-branes in AdS$_5\times$S$^5$. 
We prove that the insertion of a half-BPS Wilson loop operator in  the ${\cal N}=4$ SYM path integral is achieved by integrating out the degrees of freedom on a configuration of bulk $D5$-branes  or alternatively on a configuration of bulk   $D3$-branes. The bulk $D5$-brane and   $D3$-brane descriptions are related by bosonization.

\Date{4/2006}

\newsec{Introduction and Conclusion}

A necessary step in describing string theory in terms of a holographic dual gauge theory is to  be able to map all gauge invariant operators of the field theory in  string theory, as all physical information is captured by gauge invariant observables.

Gauge theories can be  formulated in terms of a non-abelian vector potential or alternatively in terms of  gauge invariant Wilson loop variables. The formulation in terms of non-abelian connections makes locality manifest while it has the disadvantage that the vector potential transforms inhomogeneously under  gauge transformation and is therefore not a physical observable. The formulation in terms of Wilson loop variables makes gauge invariance manifest at the expense of a lack of locality. 
The Wilson loop  variables, being non-local, appear to be the natural set of variables in which the bulk string theory formulation should be written down to make holography manifest. It is therefore interesting to consider the string theory realization of Wilson loop operators\foot{This has been done  for Wilson loops in the fundamental representation by
\lref\ReyIK{
  S.~J.~Rey and J.~T.~Yee,
  ``Macroscopic strings as heavy quarks in large N gauge theory and  anti-de
  Sitter supergravity,''
  Eur.\ Phys.\ J.\ C {\bf 22}, 379 (2001)
  [arXiv:hep-th/9803001].
}
\lref\MaldacenaIM{
  J.~M.~Maldacena,
  ``Wilson loops in large N field theories,''
  Phys.\ Rev.\ Lett.\  {\bf 80}, 4859 (1998)
  [arXiv:hep-th/9803002].
}
\ReyIK\MaldacenaIM.}.
 
 Significant progress has been made in mapping {\rm local} gauge invariant operators in gauge theory in the string theory dual. Local operators in the boundary theory correspond to bulk string fields
 \lref\MaldacenaRE{
  J.~M.~Maldacena,
  ``The large N limit of superconformal field theories and supergravity,''
  Adv.\ Theor.\ Math.\ Phys.\  {\bf 2}, 231 (1998)
  [Int.\ J.\ Theor.\ Phys.\  {\bf 38}, 1113 (1999)]
  [arXiv:hep-th/9711200].
}
\lref\WittenQJ{
  E.~Witten,
  ``Anti-de Sitter space and holography,''
  Adv.\ Theor.\ Math.\ Phys.\  {\bf 2}, 253 (1998)
  [arXiv:hep-th/9802150].
}
\lref\GubserBC{
  S.~S.~Gubser, I.~R.~Klebanov and A.~M.~Polyakov,
  ``Gauge theory correlators from non-critical string theory,''
  Phys.\ Lett.\ B {\bf 428}, 105 (1998)
  [arXiv:hep-th/9802109].
}
\lref\AharonyTI{
  O.~Aharony, S.~S.~Gubser, J.~M.~Maldacena, H.~Ooguri and Y.~Oz,
  ``Large N field theories, string theory and gravity,''
  Phys.\ Rept.\  {\bf 323}, 183 (2000)
  [arXiv:hep-th/9905111].
}
 \MaldacenaRE\WittenQJ\GubserBC\AharonyTI.
 Furthermore, the correlation function of local gauge invariant operators is obtained  by evaluating the 
 string field theory action in the bulk with prescribed sources at the boundary.
 
Wilson loop operators are an interesting set of non-local gauge invariant operators in gauge theory in which the theory can be formulated.  Mathematically, a Wilson loop is the trace in an arbitrary representation $R$ of the gauge group $G$ of the holonomy matrix associated with parallel transport along a closed curve $C$ in spacetime.  Physically, the expectation value of a Wilson loop operator in some particular representation of the gauge group measures the phase associated with moving an external charged particle with charge $R$ around a closed curve $C$ in spacetime.

In this paper we show that all half-BPS operators in four dimensional ${\cal N}=4$ SYM with gauge group $U(N)$ -- which are labeled by an irreducible representation of $U(N)$ -- 
can be  realized in the dual 
gravitational description in terms of $D5$-branes or alternatively in terms of $D3$-branes in AdS$_5\times$S$^5$. We show this by explicitly integrating out the physics on the $D5$-branes or alternatively on the $D3$-branes and proving that this inserts a half-BPS Wilson loop operator in the desired representation in the ${\cal N}=4$ SYM path integral. 
\vfill\eject
The choice of representation of $U(N)$ can be conveniently summarized in a Young tableau. We find that the data of the tableau can be precisely encoded in the AdS bulk description. Consider a Young tableau for a representation of $U(N)$  with
$n_i$ boxes in the $i$-th row  and $m_j$ boxes in the $j$-th column:
\medskip
 \ifig\Youngtabcolumnsrows{A Young tableau. For $U(N)$, $i\leq N$ and $m_j\leq N$ while $M$ and $n_i$  are arbitrary.}{\epsfxsize2in\epsfbox{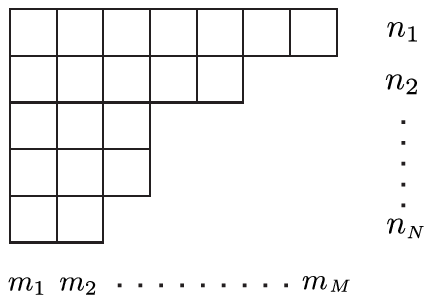}}

We show that  the Wilson operator labeled by this tableau is generated by integrating out the degrees of freedom on $M$ coincident $D5$-branes in AdS$_5\times$S$^5$ where the  $j$-th $D5$-brane has $m_j$ units of fundamental string charge dissolved in it. If we  label  the $j$-th $D5$-brane carrying $m_j$ units of charge by $D5_{m_j}$, the Young tableau in Fig. 1. has a bulk description in terms of a configuration of $D5$-branes given by $(D5_{m_1}, D5_{m_2},\ldots, D5_{m_M})$.

We show that the same Wilson loop operator can also be represented in the bulk description in terms of coincident $D3$-branes  in AdS$_5\times$S$^5$ where the  $i$-th $D3$-brane has $n_i$ units of fundamental string charge dissolved in it\foot{This $D$-brane has been  previously considered in the study of Wilson loops by Drukker and Fiol \lref\DrukkerKX{
  N.~Drukker and B.~Fiol,
  ``All-genus calculation of Wilson loops using D-branes,''
  JHEP {\bf 0502}, 010 (2005)
  [arXiv:hep-th/0501109].
}
\DrukkerKX. In this paper we show that these $D$-branes  describe  Wilson loops in a representation of the gauge group which  we determine.}. If we  label  the $i$-th $D3$-brane carrying $n_i$ units of charge by $D3_{n_i}$, the Young tableau in Fig. 1. has a bulk description in terms of a configuration of $D3$-branes\foot{The number of $D3$-branes depends on the length of the first column, which can be at most $N$. A $D3$-brane with AdS$_2\times$S$^2$ worldvolume is a domain wall in AdS$_5$ and crossing it reduces the amount of five-form flux by one unit. Having such   a $D3$-brane solution  requires the presence of five-form flux in the background to stabilize it. Therefore, we cannot put more that $N$ such $D3$-branes as inside the last one there is no more five-form flux left and the  $N+1$-th $D3$-brane cannot be stabilized.}  given by $(D3_{n_1}, D3_{n_2},\ldots, D3_{n_N})$.

 The way we show that the bulk description of half-BPS Wilson loops is given by $D$-branes is by studying the effective field theory dynamics on the $N$ $D3$-branes that generate  the AdS$_5\times$S$^5$ background in the presence of bulk $D5$ and $D3$-branes. This effective field theory  describing the coupling of the degrees of freedom on the bulk $D$-branes to the ${\cal N}=4$ SYM fields is a defect conformal field theory (see e.g
 \lref\KarchGX{
  A.~Karch and L.~Randall,
  ``Open and closed string interpretation of SUSY CFT's on branes with
  boundaries,''
  JHEP {\bf 0106}, 063 (2001)
  [arXiv:hep-th/0105132].
}
\lref\DeWolfePQ{
  O.~DeWolfe, D.~Z.~Freedman and H.~Ooguri,
  ``Holography and defect conformal field theories,''
  Phys.\ Rev.\ D {\bf 66}, 025009 (2002)
  [arXiv:hep-th/0111135].
 }
 \lref\ErdmengerEX{
  J.~Erdmenger, Z.~Guralnik and I.~Kirsch,
  ``Four-dimensional superconformal theories with interacting boundaries or
  defects,''
  Phys.\ Rev.\ D {\bf 66}, 025020 (2002)
  [arXiv:hep-th/0203020].
}
\KarchGX\DeWolfePQ\ErdmengerEX). It is by integrating out the degrees of freedom associated with the bulk $D$-branes in the defect conformal field theory that we show the correspondence between bulk branes and  Wilson loop operators. We can carry out this procedure exactly and show that this results in the insertion of a half-BPS Wilson loop operator in the ${\cal N}=4$ SYM theory and that the mapping between the Young tableau data and the bulk $D5$ and $D3$ brane configuration is the one we described above.
 
 We find that the $D3$-brane description of the Wilson loop is related to the $D5$-brane description by  bosonizing the localized degrees of freedom of the defect conformal field theory. The degrees of freedom localized in the codimension three defect, which corresponds to the location of the Wilson line, are fermions when the bulk brane is a $D5$-brane. We find that if we quantize these degrees of freedom as bosons instead, which is allowed in $0+1$ dimensions, that the defect conformal field theory captures correctly the physics of the bulk $D3$-branes. 
 
 One of outstanding issues in the gauge/gravity 
 duality is  to exhibit the origin of the loop equation of gauge theory in the gravitational description. This important problem has thus far remained elusive. Having shown that Wilson loops are more naturally described in the bulk by $D$-branes instead of by fundamental strings, it is natural to search for the origin of the   loop equation of gauge theory  in the $D$-brane picture  instead of the fundamental string picture. This is an interesting problem that we hope to turn to in the future.

 Having obtained the bulk description of all half-BPS Wilson loop operators in ${\cal N}=4$ SYM in terms of $D$-branes,  it is natural to study the Type IIB supergravity  solutions describing these Wilson loops. Precisely this program has been carried out by Lin, Lunin and Maldacena 
 \lref\LinNB{
  H.~Lin, O.~Lunin and J.~Maldacena,
  ``Bubbling AdS space and 1/2 BPS geometries,''
  JHEP {\bf 0410}, 025 (2004)
  [arXiv:hep-th/0409174].
}
\LinNB\  for the case half-BPS local operators in ${\cal N}=4$ SYM. In a recent interesting paper by Yamaguchi \YamaguchiTE, a supergravity ansatz was written down that can be used to search for these solutions. It would be interesting to solve the supergravity BPS equations for this case.

 The   description of Wilson loop operators in terms of a defect conformal field theory seems very economical and might be computationally useful when performing  calculations of correlation functions involving Wilson loops. It would also be interesting to consider the case of a circular Wilson loop\foot{Which breaks a different set of supersymmetries compared to the loops considered in this paper.}
 and study the defect field theory origin of the matrix model proposed by Erickson, Semenoff and Zarembo 
 \lref\EricksonAF{
  J.~K.~Erickson, G.~W.~Semenoff and K.~Zarembo,
  ``Wilson loops in N = 4 supersymmetric Yang-Mills theory,''
  Nucl.\ Phys.\ B {\bf 582}, 155 (2000)
  [arXiv:hep-th/0003055].
}
\lref\DrukkerRR{
  N.~Drukker and D.~J.~Gross,
  ``An exact prediction of N = 4 SUSYM theory for string theory,''
  J.\ Math.\ Phys.\  {\bf 42}, 2896 (2001)
  [arXiv:hep-th/0010274].
}
\EricksonAF\DrukkerRR\ for the study of circular Wilson loops. We expect that the description of Wilson loops studied in this paper can also be extended to other interesting gauge theories with reduced supersymmetry and different matter content.
 
 The plan of the rest of the paper is as follows. In section $2$ we identify the Wilson loop operators in  ${\cal N}=4$ SYM that preserve half of the supersymmetries  and study the ${\cal N}=4$ subalgebra  preserved by the half-BPS Wilson loops. Section $3$ contains the embeddings of the $D5_k$ and $D3_k$  brane in AdS$_5\times$S$^5$ and we show that they preserve the same symmetries as the half-BPS Wilson loop operators. In section $4$ we derive the defect conformal field theory produced by the interaction of the bulk $D5_k/D3_k$ branes with the $D3$ branes that generate the 
 AdS$_5\times$S$^5$ background. We also show that a single $D5_k$-brane corresponds to a half-BPS Wilson loop in the $k$-th antisymmetric product representation of $U(N)$ while the $D3_k$-brane
 corresponds to the  $k$-th symmetric product representation. In section $5$ we show that a half-BPS Wilson loop in any representation of $U(N)$ is described in terms of the  collection of $D5$ or $D3$ branes explained in the introduction. Some computations have been relegated to the Appendices.

 \newsec{Wilson Loops in ${\cal N}=4$ SYM}

 A Wilson loop operator in ${\cal N}=4$ SYM is labeled by a curve $C$ in superspace and by a representation $R$ of the gauge group $G$. The data that characterizes a Wilson loop, the curve $C$ and the representation $R$, label the properties of the external particle that is used to probe the theory. 
The curve $C$ 
 is identified with the worldline of the  superparticle propagating in ${\cal N}=4$ superspace while the representation $R$ corresponds to the charge carried by the superparticle. 
 
 The curve $C$ is parameterized by $(x^\mu(s),y^I(s),\theta^\alpha_A(s))$
 and it encodes the coupling of the charged external  superparticle to the 
 ${\cal N}=4$ SYM multiplet $(A_\mu, \phi^I,\lambda_\alpha^A)$, where $\mu\ (\alpha)$ is a vector(spinor) index of $SO(1,3)$ while $I\ (A)$ is a vector (spinor) index of the $SO(6)$ R-symmetry group of ${\cal N}=4$ SYM. Gauge invariance of the Wilson loop constraints the curve $x^\mu(s)$ to be closed while $(y^I(s),\theta^\alpha_A(s))$ are arbitrary curves.

The other piece of data entering into the definition  of a Wilson loop operator is the choice of representation $R$ of the gauge group $G$.  For gauge group $U(N)$, the irreducible representations are conveniently summarized by a Young tableau $R=(n_1,n_2,\ldots,n_N)$, where $n_i$ is the number of boxes in the $i$-th row of the tableau and $n_1\geq n_2\geq \ldots\geq n_N\geq 0$. 
The corresponding Young diagram is given by:

\medskip

\centerline{\young(12\cdot\cdot\cdot\cdot\none,12\cdot\cdot\cdot\ntwo,12\cdot\cdot\cdot\nthree,\cdot\cdot\cdot\cdot,12\cdot\nen)}
\medskip
\noindent
The main goal of this paper is to identify  all half-BPS Wilson loop operators of ${\cal N}=4 $ SYM in the dual asymptotically AdS gravitational description.

In this paper we consider bosonic Wilson loop operators for which $\theta^\alpha_A(s)=0$. Wilson loop operators coupling to fermions can be obtained by the action of supersymmetry and are descendant operators. The operators under study are given by
 \eqn\Wilson{
 W_R(C)=\hbox{Tr}_{R}\;P \exp\left(i\int_C ds (A_\mu\dot{x}^\mu+\phi_I \dot{y}^I)\right),}
 where $C$ labels the curve $(x^\mu(s),y^I(s))$ and $P$ denotes path-ordering along the curve $C$.

We now consider the Wilson loop operators in ${\cal N}=4$ SYM which are invariant under one-half of the ${\cal N}=4$ Poincare supersymmetries  and also invariant under one-half of the ${\cal N}=4$ superconformal  supersymmetries. The sixteen Poincare supersymmetries are generated by a ten dimensional Majorana-Weyl spinor $\epsilon_1$ of negative chirality while the superconformal  supersymmetries are generated by a ten dimensional Majorana-Weyl spinor $\epsilon_2$ of positive chirality. The analysis in Appendix A  shows that supersymmetry restricts the curve $C$ to be a straight time-like line spanned by $x^0=t$ and $\dot y^I=n^I$, where $n^I$ is a unit vector in $R^6$. The unbroken supersymmetries are generated by $\epsilon_{1,2}$ satisfying 
\eqn\unbrokensusy{
\gamma_0\gamma_I n^I\epsilon_1=\epsilon_1\qquad \gamma_0\gamma_I n^I\epsilon_2=-\epsilon_2.}

 Therefore, the half-BPS Wilson loop operators in ${\cal N}=4$ SYM are given by
\eqn\Wilsonsusy{
 W_R=W_{(n_1,n_2,\ldots,n_N)}=\hbox{Tr}_{R}\;P \exp\left(i \int dt\; (A_0+\phi)\right),}
 where $\phi=\phi_In^I$. It follows that the half-BPS Wilson loop operators carry only one label:  the choice of representation $R$.

We conclude this section by exhibiting the supersymmetry algebra preserved by the insertion of \Wilsonsusy\ to the ${\cal N}=4$ path integral. This becomes useful  when identifying the gravitational dual description of  Wilson loops in later sections.    In the absence of  any operator insertions,  ${\cal N}=4$ SYM is invariant under the $PSU(2,2|4)$ symmetry group. It is well known
\lref\CardyBB{
  J.~L.~Cardy,
  ``Conformal Invariance And Surface Critical Behavior,''
  Nucl.\ Phys.\ B {\bf 240}, 514 (1984).
} 
\CardyBB\
that a straight line breaks the four dimensional conformal group  $SU(2,2)\simeq SO(2,4)$ down to $SO(4^*)\simeq SU(1,1)\times SU(2)\simeq SL(2,R)\times SU(2)$. Moreover, the choice of a unit vector $n^I$ in \Wilsonsusy\ breaks the $SU(4)\simeq SO(6)$ R-symmetry of ${\cal N}=4$ SYM down to  $Sp(4)\simeq SO(5)$. 
The projections \unbrokensusy\ impose a reality condition on the four dimensional supersymmetry generators, which now transform in the $(4,4)$ representation of $SO(4^*)\times Sp(4)$. Therefore, the supersymmetry algebra preserved\foot{This supergroup has appeared in the past in relation to the baryon vertex
\lref\CrapsNC{
  B.~Craps, J.~Gomis, D.~Mateos and A.~Van Proeyen,
  ``BPS solutions of a D5-brane world volume in a D3-brane background from
  superalgebras,''
  JHEP {\bf 9904}, 004 (1999)
  [arXiv:hep-th/9901060].
}
\lref\GomisXG{
  J.~Gomis, P.~K.~Townsend and M.~N.~R.~Wohlfarth,
  ``The 's-rule' exclusion principle and vacuum interpolation in worldvolume
  dynamics,''
  JHEP {\bf 0212}, 027 (2002)
  [arXiv:hep-th/0211020].
}
\CrapsNC\GomisXG.}
by the half-BPS Wilson loops is $Osp(4^*|4)$.

\newsec{Giant and Dual Giant Wilson loops} 

The goal of this section is to put forward plausible candidate $D$-branes for the bulk description of the  half-BPS Wilson loop operators \Wilsonsusy. In the following sections we show that integrating out the physics on these D-branes results in the insertion of  a half-BPS Wilson loop operator to ${\cal N}=4$ SYM. This provides the string theory realization of all half-BPS Wilson loops in ${\cal N}=4$ SYM.

Given the extended nature of Wilson loop operators in  the gauge theory living at the boundary of AdS, it is natural to search for extended objects in AdS$_5\times$S$^5$ preserving  the same symmetries as those preserved by  the half-BPS operators \Wilsonsusy\ as candidates for the  dual description of Wilson loops. The extended objects that couple to the Wilson loop must be such that they span a time-like line in the boundary of AdS, where the Wilson loop operator \Wilsonsusy\ is defined.  

Since we want to identify extended objects with Wilson loops in
 ${\cal N}=4$ SYM on $R^{1,3}$, it is convenient to write the AdS$_5$ metric in Poincare coordinates
\eqn\ads{
ds^2_{AdS}=L^2\left(
u^2\eta_{\mu\nu}dx^\mu dx^\nu+{du^2\over u^2}\right),}
where $L=(4\pi g_s N)^{1/4}l_s$ is the radius of AdS$_5$ and S$^5$.
Furthermore, since the Wilson loop operator \Wilsonsusy\ preserves an $SO(5)$ symmetry, we make this symmetry manifest by foliating  the metric on S$^5$ by a family of S$^4$'s
\eqn\sphere{
ds^2_{sphere}=L^2\left(d\theta^2+\sin^2\theta\; d\Omega_4^2\right),}
where $\theta$ measures the latitude angle of the S$^4$ from the north pole and $d\Omega_4^2$ is the metric on the unit S$^4$.

In \ReyIK\MaldacenaIM\  the bulk description of a Wilson loop in the fundamental representation of the gauge group associated with a curve $C$ in $R^{1,3}$   was given in terms of  a fundamental string propagating in the bulk and ending at the boundary of AdS along the curve $C$. This case corresponds to the simplest Young tableau $R=(1,0,\ldots,0)$, with Young diagram $\yng(1)$.

The expectation value of the corresponding Wilson loop operator is identified with the action of the string ending at the boundary along $C$. This identification was motivated by considering a stack of D3-branes and moving one of them to infinity, leaving behind a massive external particle carrying charge in the fundamental representation of the gauge group.

The embedding corresponding to the half-BPS Wilson loop \Wilsonsusy\  for $R=(1,0,\ldots,0)$ is  given by\foot{The coordinates $\sigma^0,\ldots \sigma^p$ refer to the worldvolume coordinates on a string/brane.}
\eqn\embedd{
\sigma^0=x^0\qquad \sigma^1=u       \qquad x^i=0\qquad x^I=n^I,}
so that the fundamental string spans an AdS$_2$ geometry sitting at  $x^i=0$ in AdS$_5$  and sits at a point on the S$^5$ labeled by a unit vector $n^I$, satisfying $n^2=1$.  Therefore, the fundamental string preserves exactly the same $SU(1,1)\times SU(2)\times SO(5)$ symmetries as the one-half BPS Wilson loop operator \Wilsonsusy. Moreover the string ends on the time-like line parameretrized by $x^0=t$, which is the curve corresponding to the half-BPS Wilson loop \Wilsonsusy. 

In Appendix $B$ we compute the supersymmetries left unbroken by the fundamental string \embedd. We find that they are generated by two ten dimensional Majorana-Weyl spinors $\epsilon_{1,2}$ of opposite chirality satisfying 
\eqn\project{
\gamma_0\gamma_In^I\epsilon_1=\epsilon_1\qquad \gamma_0\gamma_In^I\epsilon_2=-\epsilon_2,}
which coincides with the unbroken supersymmetries \unbrokensusy\ of the half-BPS Wilson loop. Therefore, the fundamental string preserves the same $Osp(4^*|4)$ symmetry as the half-BPS Wilson loop \Wilsonsusy. 

The main question in this paper is, what is the holographic  description of half-BPS Wilson loop operators in higher representations of the gauge group?

 Intuitively, higher representations correspond to having multiple coincident fundamental strings\foot{Such a proposal was put forward in 
 \lref\GrossGK{
  D.~J.~Gross and H.~Ooguri,
  ``Aspects of large N gauge theory dynamics as seen by string theory,''
  Phys.\ Rev.\ D {\bf 58}, 106002 (1998)
  [arXiv:hep-th/9805129].
}
\GrossGK\ by drawing lessons from the description of Wilson loops in two dimensonal QCD.}  ending at the boundary of AdS. This description is, however, not very useful as the Nambu-Goto action only describes a single string. A better description of the system is achieved by realizing that   coincident fundamental strings in the AdS$_5\times$S$^5$ background can polarize
 \lref\MyersPS{
  R.~C.~Myers,
  ``Dielectric-branes,''
  JHEP {\bf 9912}, 022 (1999)
  [arXiv:hep-th/9910053].
}
\MyersPS\ into a single D-brane with fundamental strings dissolved in it, thus providing a concrete description of the coincident fundamental strings. 

We now describe the way 
in which a collection of $k$ fundamental strings puff up into a D-brane with  $k$ units of fundamental string charge on the D-brane worldvolume. 

The guide we use to determine which D-branes are the puffed up description 
 of $k$-fundamental strings is to consider D-branes in AdS$_5\times$S$^5$ which are invariant under the same symmetries as the half-BPS Wilson loops\foot{We have already established that the fundamental strings \embedd\ have the same symmetries as the half-BPS Wilson loops.}, namely we demand invariance under $Osp(4^*|4)$. 
The branes preserving the $SU(1,1)\times SU(2)\times SO(5)$ symmetries of the Wilson loop are
 given by:

\noindent
1) $D5_k$-brane with AdS$_2\times$S$^4$ worldvolume.

\noindent
2) $D3_k$-brane with AdS$_2\times$S$^2$ worldvolume. 

We now describe the basic properties of these branes that we need for the  analysis in upcoming sections.

\subsec{$D5_k$-brane as a  Giant Wilson loop} 

The classical equations of motion for a $D5$-brane with an AdS$_2\times$S$^4$  geometry and with $k$ fundamental strings dissolved in it (which we label by $D5_k$) has been studied in the past in 
\lref\PawelczykHY{
  J.~Pawelczyk and S.~J.~Rey,
  ``Ramond-Ramond flux stabilization of D-branes,''
  Phys.\ Lett.\ B {\bf 493}, 395 (2000)
  [arXiv:hep-th/0007154].
}
\lref\CaminoAT{
  J.~M.~Camino, A.~Paredes and A.~V.~Ramallo,
  ``Stable wrapped branes,''
  JHEP {\bf 0105}, 011 (2001)
  [arXiv:hep-th/0104082].
}
\PawelczykHY\CaminoAT. Here we summarize the necessary elements that will allow us to prove in the following section that this D-brane corresponds to a half-BPS Wilson loop operator. 

The $D5_k$-brane is described by the following embedding
\eqn\embedddfive{
\sigma^0=x^0\qquad \sigma^1=u \qquad \sigma^a=\varphi_a\qquad x^i=0\qquad \theta=\theta_k=\hbox{constant},} 
together with a nontrivial electric field $F$  along the AdS$_2$  spanned by $(x^0,u)$. Therefore, a  $D5_k$-brane spans an AdS$_2\times$S$^4$ geometry\foot{ $\varphi_a$ are the coordinates on the S$^4$ in \sphere.} and  sits at a latitude angle $\theta=\theta_k$ on the S$^5$, which depends on $k$, the fundamental string charge carried by the $D5_k$-brane:

\ifig\sphera{A $D5_k$-brane sits at a latitude angle $\theta_k$  determined by the amount of fundamental string charge it carries.}{\epsfxsize1.5in\epsfbox{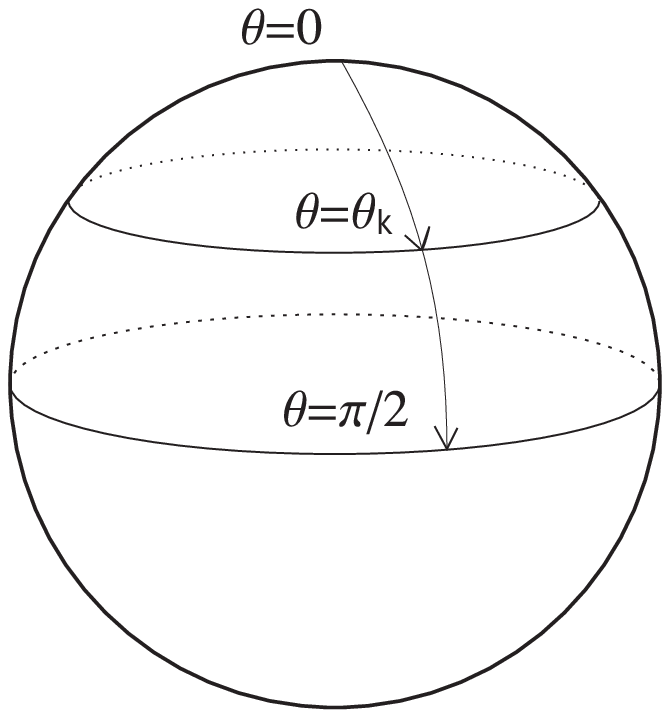}}

This brane describes the puffing up of $k$ fundamental strings into a D-brane inside S$^5$, so in analogy with a similar phenomenon for point-like gravitons 
\lref\McGreevyCW{
  J.~McGreevy, L.~Susskind and N.~Toumbas,
  ``Invasion of the giant gravitons from anti-de Sitter space,''
  JHEP {\bf 0006}, 008 (2000)
  [arXiv:hep-th/0003075].
}
\McGreevyCW, such a brane can be called a giant Wilson loop.

It can be shown \CaminoAT\  that $\theta_k$  is a monotonically increasing function of $k$ in the domain of $\theta$, that is $[0,\pi]$ and that $\theta_0=0$ and $\theta_N=\pi$, where $N$ is the amount of flux in the AdS$_5\times$S$^5$ background or equivalently the rank of the gauge group in ${\cal N}=4$ SYM. Therefore, we can dissolve at most $N$ fundamental strings on the $D5$-brane.

 The $D5_k$-brane has  the same bosonic symmetries as the half-BPS Wilson loop operator and it ends on the boundary of AdS$_5$ along the time-like line where the half-BPS Wilson loop operator \Wilsonsusy\ is defined. In Appendix $B$ we show that it also preserves the same supersymmetries \unbrokensusy\ as the half-BPS Wilson loop operator \Wilsonsusy\ when $n^I=(1,0,\ldots,0)$ and is therefore invariant under the $Osp(4^*|4)$ symmetry group.

\subsec{$D3_k$-brane as a  Dual Giant Wilson loop} 

The classical equations of motion of a $D3$-brane with an AdS$_2\times$S$^2$ geometry   and with $k$ fundamental strings dissolved in it (which we label by $D3_k$) has been studied recently by Drukker and Fiol \DrukkerKX. We refer the reader to this reference for the details of the solution.

 For our purposes we note that unlike for the case of the $D5_k$-brane, an arbitrary amount of fundamental string charge can be dissolved on the $D3_k$-brane. As we shall see in the next section, this  has a pleasing interpretation in   ${\cal N}=4$. 
 
 The geometry spanned by  a $D3_k$-brane gives an    AdS$_2\times$S$^2$  foliation\foot{This foliation structure and the relation with ${\cal N}=4$ SYM defined on the AdS$_2\times$S$^2$  boundary -- which makes manifest the symmetries left unbroken by the insertion of a straight line Wilson loop -- has been considered in 
\lref\KapustinPY{
  A.~Kapustin,
  ``Wilson-'t Hooft operators in four-dimensional gauge theories and
  S-duality,''
  arXiv:hep-th/0501015.
}
\lref\GomisPG{
  J.~Gomis, J.~Gomis and K.~Kamimura,
  ``Non-relativistic superstrings: A new soluble sector of AdS(5) x S**5,''
  JHEP {\bf 0512}, 024 (2005)
  [arXiv:hep-th/0507036].
}
\KapustinPY\GomisPG\YamaguchiTE.}
  of AdS$_5$, the location of the slice being determined by $k$, the amount of fundamental string charge:

 \ifig\sphera{A $D3_k$-brane gives an AdS$_2\times$S$^2$ slicing of AdS$_5$.}{\epsfxsize1.5in\epsfbox{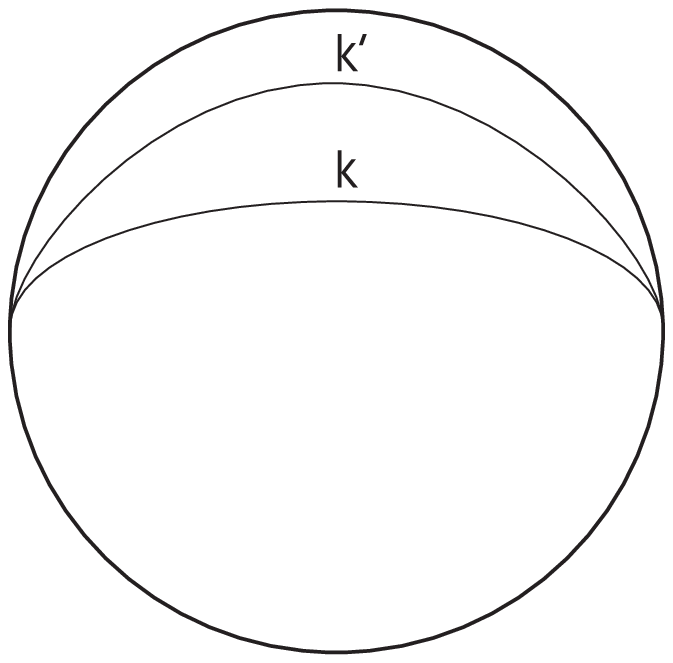}}  
 This brane describes the puffing up of $k$ fundamental strings into a D-brane inside AdS$_5$, so in analogy with a similar phenomenon for point-like gravitons 
 \lref\GrisaruZN{
  M.~T.~Grisaru, R.~C.~Myers and O.~Tafjord,
  ``SUSY and Goliath,''
  JHEP {\bf 0008}, 040 (2000)
  [arXiv:hep-th/0008015].
}
\lref\HashimotoZP{
  A.~Hashimoto, S.~Hirano and N.~Itzhaki,
  ``Large branes in AdS and their field theory dual,''
  JHEP {\bf 0008}, 051 (2000)
  [arXiv:hep-th/0008016].
} 
 \GrisaruZN\HashimotoZP, such a brane can be called a dual giant Wilson loop.

By generalizing the supersymmetry analysis in \DrukkerKX\ one can show that the $D3_k$-brane preserves precisely the same supersymmetries as the fundamental string
\project\ and therefore the same as the ones preserved by the half-BPS Wilson loop operator.

To summarize, we have seen that $k$ fundamental strings can be described either by a single $D5_k$-brane or by a single $D3_k$-brane. The three objects preserve the same $Osp(4^*|4)$ symmetry if the fundamental string and the $D3_k$-brane sit at the north pole of the S$^5$, i.e. at $\theta=0$ corresponding to the unit vector $n^I=(1,0,\ldots,0)$. Furthermore, these three objects are invariant under the same $Osp(4^*|4)$ symmetry  as the half-BPS Wilson loop operator \Wilsonsusy .

\newsec{Dirichlet Branes  as  Wilson loops}

We show that the half-BPS Wilson loop operators in ${\cal N}=4$ SYM are realized by the $D$-branes in the previous section. We 
study the modification on  the low energy effective field theory 
on the $N$ $D$3-branes that generate the AdS$_5\times$S$^5$ background due to the presence of $D5$-brane giants and $D3$-brane dual giants. We can integrate out exactly the degrees of freedom introduced by the Wilson loop $D$-branes and  show that the net effect of these $D$-branes is to insert into the ${\cal N}=4$ $U(N)$ SYM path integral a Wilson loop operator in the desired representation of the $U(N)$ gauge group.

In order to develop some intuition for how this procedure works, we start by analyzing  the case of a single $D5_k$-brane and  a single $D3_k$-brane. We  now show that a $D5_k$-brane describes a half-BPS Wilson loop operator in the $k$-th antisymmetric product representation of $U(N)$ while a $D3_k$-brane describes one in the $k$-th symmetric product representation.

In section $5$ we proceed to show that a Wilson loop described by  an arbitrary Young tableau corresponds to considering multiple $D$-branes. We also show that a  given Young tableau can be either derived from a collection of $D5_k$-branes or from a collection of $D3_k$-branes and  that the two descriptions are related by bosonization.

\subsec{$D5_k$-brane  as a Wilson Loop}

We propose to analyze the physical interpretation 
of a single $D5_k$-brane in the gauge theory by studying the effect it has on four dimensional ${\cal N}=4$ SYM.  A $D5_k$-brane with an AdS$_2\times$S$^4$ worldvolume in AdS$_5\times$S$^5$ arises in the  near horizon limit of a single $D5$-brane probing the $N$ $D3$-branes that generate the AdS$_5\times$S$^5$ background. The flat space brane configuration is given by:
\eqn\braneconf{\matrix{\ \ &0&1&2&3&4&5&6&7&8&9\cr
D3&\hbox{X}&\hbox{X}&\hbox{X}&\hbox{X}&&&&&\cr
D5&\hbox{X}&&&&&\hbox{X}&\hbox{X}&\hbox{X}&\hbox{X}&\hbox{X}}}
\medskip

\noindent
We can now study the effect of the $D5_k$-brane by analyzing the low energy effective field theory on a single $D5$-brane probing $N$ $D3$-branes in flat space 

We note first that the $D5$-brane produces  a codimension three defect on the $D3$-branes, since they overlap only in the time direction. In order to derive the decoupled field theory we must analyze the various open string sectors. The 3-3 strings give rise to the the familiar  four dimensional ${\cal N}=4$ $U(N)$ SYM theory. The sector of 3-5 and 5-3 strings give rise to degrees of freedom that are localized in the defect. There are also the 5-5 strings. The degrees of freedom associated with these strings --  a six dimensional  vector multiplet on the $D5$-brane -- are not dynamical. Nevertheless, as we will see, they play a crucial role in encoding the choice of Young tableau $R=(n_1,\ldots,n_N)$.

This brane configuration gives rise to a defect conformal field theory
(see e.g.  \KarchGX\DeWolfePQ), which describes the coupling of the ${\cal N}=4$ SYM to the localized degrees of freedom. The localized degrees of freedom arise from the 3-5 and 5-3 strings and they give rise to  fermionic fields $\chi$ transforming in the fundamental representation of $U(N)$. We can  write the action of  this defect conformal field theory by realizing that we can obtain it by performing T-duality on the well studied D0-D8 matrix quantum mechanics (see  e.g.   
\lref\BanksZS{
  T.~Banks, N.~Seiberg and E.~Silverstein,
  ``Zero and one-dimensional probes with N = 8 supersymmetry,''
  Phys.\ Lett.\ B {\bf 401}, 30 (1997)
  [arXiv:hep-th/9703052].
}
 \lref\BachasKN{
  C.~P.~Bachas, M.~B.~Green and A.~Schwimmer,
  ``(8,0) quantum mechanics and symmetry enhancement in type I'
  superstrings,''
  JHEP {\bf 9801}, 006 (1998)
  [arXiv:hep-th/9712086].
}
\BanksZS\BachasKN). Ignoring for the moment the coupling of $\chi$ to the non-dynamical 5-5 strings, we obtain that the action of our defect conformal field theory is given by\foot{We do not write the $U(N)$ indices explicitly. They are contracted in a straighforward manner between the $\chi_i$ fields and the $A_{0\;ij}$ gauge field, where $i,j=1,\ldots,N$.}
\eqn\lag{
S=S_{{\cal N}=4}+\int dt\;i \chi^\dagger \partial_t \chi+\chi^\dagger(A_0+\phi)\chi,}
where $A_0$ is the temporal component of the gauge field in ${\cal N}=4$ SYM and $\phi$ is one of the scalars of ${\cal N}=4$ SYM describing the position of the $D3$-branes in the direction transverse to both the $D3$ and $D5$ branes;  it corresponds to the unit vector $n^I=(1,0,\ldots,0)$.

What are the $PSU(2,2|4)$ symmetries   that are left unbroken by adding to the ${\cal N}=4$ action the localized fields? The supersymmetries of ${\cal N}=4$ SYM act trivially on $\chi$. This implies that the computation determining  the unbroken supersymmetries is exactly the same as the one we did for the Wilson loop operator \Wilsonsusy. Likewise for  the bosonic symmetries, where we just need to note that the defect fields live  on a time-like straight line. Therefore, we conclude that our defect conformal field theory has an $Osp(4^*|4)$ symmetry, just like the half-BPS Wilson loop operator \Wilsonsusy.

Even though the fields arising from the 5-5 strings are nondynamical, they play a crucial  role in the identification of the $D5_k$-brane with a Wilson loop operator  in a particular representation of the gauge group.
As we discussed in the previous section, a $D5_k$-brane has $k$ fundamental strings ending on it and we must find a way to encode the choice of $k$ in the low energy effective field theory on the $D$-branes in flat space. This can be accomplished by recalling that a fundamental string ending on a $D$-brane behaves as an electric charge for the gauge field living on the $D$-brane. Therefore we must add to \lag\ a term that captures the fact that there are $k$ units of background electric charge localized on the defect. This is accomplished by inserting into our defect conformal field theory path integral the operator:
\eqn\insercharge{
\exp\left(-i k \int  dt\; \tilde{A}_0\right).}
Equivalently, we must add to the action \lag\  the  Chern-Simons term:
\eqn\cherncharge{
-  \int dt\; k \tilde{A}_0.}
The effect of \cherncharge\ on the $\tilde{A}_0$ equation of motion is to insert $k$ units of electric charge at the location of the defect, just as desired. 

 We must also consider the coupling of the 
$\chi$ fields to the nondynamical gauge field $\tilde{A}$ on the $D5$-brane, as they transform in the fundamental representation of the $D5$-brane gauge field. 
Summarizing, we must add to \lag\ :
\eqn\extra{
S_{extra}=\int dt\; \chi^\dagger \tilde{A}_0\chi-k \tilde{A}_0.}
The addition of these extra couplings preserves the $Osp(4^*|4)$ symmetry of our defect conformal field theory.

We want to  prove that a $D5_k$-brane  corresponds to a half-BPS  Wilson loop operator in  ${\cal N}=4$ SYM in  a very specific representation of $U(N)$. The way we show this is by integrating out explicitly the degrees of freedom associated with the $D5_k$-brane. We must calculate the following path integral
\eqn\path{
Z=\int [D\chi ] [D\chi^\dagger] [D\tilde{A}_0] \; e^{i(S+S_{extra})},}
where $S$ is given in \lag\ and $S_{extra}$ in \extra.

Let's us ignore the effect of $S_{extra}$ for the time being;  we will take it into account later. We first  integrate out the $\chi$  fields.  This can be accomplished the easiest by perfoming a choice of gauge  such that the matrix $A_0+\phi$ has constant eigenvalues\foot{Here there is a subtlety. This gauge choice introduces a Fadeev-Popov determinant which changes the measure of the path-integral over the ${\cal N}=4$ SYM fields. Nevertheless, after we integrate out the degrees of freedom associated with the $D5$-brane, we can write the result in a gauge invariant form, so that the Fadeev-Popov determinant can be reabsorbed to yield the usual measure over the ${\cal N}=4$ SYM fields in the path integral.}:
\eqn\diagon{
A_0+\phi=\hbox{diag}(w_1,\ldots,w_N).}
The equations  of motion for the $\chi$ fields  are then given by:
\eqn\motion{
(i\partial_t+w_i)\chi_i=0\qquad \hbox{for} \qquad i=1,\ldots,N.}
Therefore, in this gauge, one has a system of $N$ fermions  $\chi_i$ with energy $w_i$. 

The path integral can now be conveniently evaluated by going to the Hamiltonian formulation, where integrating out the $\chi$ fermions corresponds to evaluating the partition function of the fermions\foot{Here we introduce, for convenience an infrared  regulator, so that $t$ is compact $0\leq t\leq \beta$.}.  Therefore, we are left with
\eqn\out{
Z^*=e^{iS_{{\cal N}=4}}\cdot \prod_{i=1}^N(1+x_i),}
where $x_i=e^{i \beta w_i}$ and the  $*$ in \out\ is to remind us that we have not yet taken into account the effect of $S_{extra}$ in \path.  A first glimpse of the connection between a $D5_k$-brane and a half-BPS Wilson loop operator is to recognize that the quantity $x_i=e^{i \beta w_i}$ appearing in \out\ with $w_i$ given in \diagon,  is an  eigenvalue of the holonomy matrix appearing in the Wilson loop operator \Wilsonsusy, that is $\exp i\beta \left(A_0+\phi \right)$.

Since our original path integral \path\  is invariant under $U(N)$ conjugations, it means that $Z^*$ should have an expansion in terms of characters or invariant traces  of $U(N)$, which are labeled by a Young tableau $R=(n_1,n_2,\ldots,n_N)$.  In order to exhibit which representations $R$ appear in the partition function,  we split the computation of the partition function into sectors with a fixed number of fermions in a state. This decomposition allows us to write 
\eqn\expand{
\prod_{i=1}^N(1+x_i)=\sum_{l=0}^{N}E_{l}(x_1,\ldots,x_l),}
where $E_{l}(x_1,\ldots,x_l)$ is the symmetric polynomial:
\eqn\poly{
E_{l}(x_1,\ldots,x_l)=\sum_{i_1<i_2\ldots <i_l}x_{i_1}\ldots x_{i_l}.}
Physically, $E_{l}(x_1,\ldots,x_l)$ is the partition function over the Fock space
of $N$ fermions, each with  energy $w_i$,   that have $l$ fermions in a state.

We now recognize that the polynomial $E_l$ is the formula (see e.g
\lref\groupthe{
W. Fulton and J. Harris, ``Representation Theory", Springer 2000.
  }
\groupthe)
for the trace of the half-BPS Wilson loop holonomy matrix in the $l$-th antisymmetric representation
\eqn\antis{
E_l= \hbox{Tr}_{\matrix{(\underbrace{1, \ldots ,1},0,\ldots ,0)\cr
\hskip-34pt l}}\;P \exp\left(i \int dt\; (A_0+\phi)\right)=W_{\matrix{(\underbrace{1, \ldots ,1},0,\ldots ,0)\cr
\hskip-34pt l}},}
where $W_{\matrix{(\underbrace{1, \ldots ,1},0,\ldots ,0)\cr
\hskip-34pt l}}$
 is the half-BPS Wilson loop operator \Wilsonsusy\ corresponding to the following Young diagram:

\noindent
\centerline{\young(1,2,\cdot,\cdot,\cdot,l)
}
Therefore, integrating out the $\chi$ fields has the effect of  inserting into the ${\cal N}=4$ path integral a sum over all half-BPS Wilson loops in the $l$-th antisymmetric representation:
\eqn\outa{
Z^*=e^{iS_{{\cal N}=4}}\cdot \sum_{l=0}^{N}W_{\matrix{(\underbrace{1, \ldots ,1},0,\ldots ,0)\cr
\hskip-34pt l}}.}

It is  now easy to go back and consider  the effect of  $S_{extra}$ \extra\ on the path integral   \path. Integrating over $\tilde{A}_0$ in \path\ imposes the following constraint:
\eqn\constraint{
\sum_{i=1}^N\chi_i^\dagger\chi_i=k.}
This constraint restrict the sum over states in  the partition function to  states with precisely $k$ fermionic excitations. These states are of the form:
\eqn\states{
\chi_{i_1}^\dagger \ldots \chi_{i_k}^\dagger|0\rangle.}
This picks out the term with $l=k$ in \outa . 

Therefore, we have shown that a single $D5_k$-brane inserts a half-BPS operator in the $k$-th antisymmetric representation in the ${\cal N}=4$ path integral
\eqn\pathfin{
D5_k\longleftrightarrow Z=e^{iS_{{\cal N}=4}}\cdot W_{\matrix{(\underbrace{1, \ldots ,1},0,\ldots ,0)\cr
\hskip-34pt k}},}
where $S_{{\cal N}=4}$ is the action of ${\cal N}=4$ SYM.
The expectation value of this operator can be computed by evaluating the classical action of the $D5_k$-brane.

\subsec{$D3_k$-brane as a Wilson Loop}

We now consider what a $D3_k$ dual giant brane corresponds to in  four dimensional ${\cal N}=4$ 
SYM.
Here we run into a puzzle. Unlike for the case of a  $D5_k$-brane, where we could study the physics produced by the brane
by identifying a brane configuration in flat space that gives rise to a  $D5_k$-brane in AdS$_5\times$S$^5$ in the near horizon/decoupling limit, there is no brane configuration in flat space that gives rise  in the near horizon/decoupling limit to a $D3_k$-brane\foot{{\bf Note Added}. The  flat space brane configuration  that reproduces the $D3_k$-brane solution in the near horizon limit has been discussed in 
\lref\GomisIM{
  J.~Gomis and F.~Passerini,
  ``Wilson loops as D3-branes,''
  arXiv:hep-th/0612022.
}  \GomisIM. } in 
AdS$_5\times$S$^5$.

Despite this shortcoming we make a very simple proposal for how to study the effect produced by a  $D3_k$-brane on ${\cal N}=4$ SYM and show that it leads to a consistent physical picture. The basic observation is that if we quantize the $\chi$ fields appearing in \lag\extra\ not as fermions but as bosons, which is something that is consistent when quantizing degrees of freedom in $0+1$ dimensions,  we can show that the effect of the $D3_k$-brane is to insert a half-BPS Wilson loop operator \Wilsonsusy\ in the $k$-th symmetric representation of $U(N)$.

This result is in concordance with the basic physics of the probe branes.  In the previous section we found that the amount of fundamental string charge $k$ on a $D5_k$-brane can be at most $N$. On the other hand, we have shown that a $D5_k$-brane corresponds to a Wilson loop in the $k$-th antisymmetric representation of $U(N)$ so that indeed $k\leq N$, otherwise the operator vanishes.  
For the $D3_k$-brane, however, the string charge $k$ can be made  arbitrarily large. The proposal  that the $D3_k$-brane can be studied in the gauge theory by quantizing $\chi$ as bosons leads, as we will show, to a Wilson loop in the $k$-th symmetric representation, for which there is a non-trivial representation of $U(N)$ for all $k$ and fits nicely with the $D3_k$-brane probe expectations.

Formally, going from the $D5_k$ giant to the $D3_k$ dual  giant Wilson line picture amounts to performing a bosonization of the defect field $\chi$ . It would be very interesting to understand from a more microscopic perspective the origin of this bosonization\foot{A similar type of bosonization seems to be at play in the description of half-BPS local operators in ${\cal N}=4$ SYM in terms of giants and dual giant gravitons
\lref\DharFG{
  A.~Dhar, G.~Mandal and N.~V.~Suryanarayana,
  ``Exact operator bosonization of finite number of fermions in one space
  dimension,''
  JHEP {\bf 0601}, 118 (2006)
  [arXiv:hep-th/0509164].
}
\DharFG.}.

Having motivated treating $\chi$ as a boson we can now go ahead and integrate out the $\chi$ fields in \path. As before, we ignore for the time being the effect of $S_{extra}$ in \path. We also diagonalize the matrix $A_0+\phi$ as in \diagon. 

The equations of motion are now those for $N$ chiral bosons $\chi_i$ with energy  $w_i$
\eqn\motionbose{
(i\partial_t+w_i)\chi_i=0\qquad \hbox{for} \qquad i=1,\ldots,N,}
where $w_i$ are the eigenvalues of the matrix $A_0+\phi$.

The path integral over $\chi$ in \path\ is computed by evaluating the partition function of the chiral bosons, which yield
\eqn\outbose{
Z^*=e^{iS_{{\cal N}=4}}\cdot \prod_{i=1}^N{1\over 1-x_i},}
where $x_i=e^{i \beta w_i}$ and the  $*$ in \out\ is to remind us that we have not yet taken into account the effect of $S_{extra}$ in \path. $x_i$ are the  eigenvalues of the holonomy matrix appearing in the Wilson loop operator \Wilsonsusy.

In order to connect this computation with Wilson loops in ${\cal N}=4$ SYM 
it is convenient to decompose the Fock space of the chiral bosons in terms of subspaces with a fixed number of bosons in a state. This decomposition yields
\eqn\expandbose{
\prod_{i=1}^N{1\over 1-x_i}=\sum_{l=0}^{\infty}H_l(x_1,\ldots,x_l), }
where $H_l(x_1,\ldots,x_l)$ is the symmetric polynomial:
\eqn\polybose{
H_{l}(x_1,\ldots,x_l)=\sum_{i_1\leq i_2\ldots \leq i_l}x_{i_1}\ldots x_{i_l}.}
Physically, $H_{l}(x_1,\ldots,x_l)$ is the partition function over the Fock space
of $N$ chiral bosons with energy $w_i$  that have $l$ bosons in a state.

We now recognize that the polynomial $H_l$ is the formula
 (see e.g
\groupthe)
for the trace of the half-BPS Wilson loop holonomy matrix in the $l$-th symmetric representation
\eqn\symm{
H_l= \hbox{Tr}_{(l,0, \ldots ,0)}\;P \exp\left(i \int dt\; (A_0+\phi)\right)=W_{(l,0, \ldots ,0)},}
where $W_{(l,0, \ldots ,0)}$
 is the half-BPS Wilson loop operator \Wilsonsusy\ corresponding to the following Young diagram:
\medskip
\noindent
\centerline{\young(12\cdot\cdot\cdot\cdot l)
}
\medskip

Therefore, integrating out the $\chi$ fields has the effect of  inserting into the ${\cal N}=4$ path integral a sum over all half-BPS Wilson loops in the $l$-th symmetric representation:
\eqn\outa{
Z^*=e^{iS_{{\cal N}=4}}\cdot \sum_{l=0}^{N}W_{(l,0, \ldots ,0)}.}

It is now straightforward to take into account the effect of $S_{extra}$ \extra\ in \path. Integrating over $\tilde{A}_0$ imposes the constraint \constraint . This constraint picks out states with precisely $k$ bosons \states\ and therefore selects the term with $l=k$ in \expandbose . 

Therefore, we have shown that a single $D3_k$-brane inserts a half-BPS operator in the $k$-th symmetric representation in the ${\cal N}=4$ path integral
\eqn\pathfin{
D3_k\longleftrightarrow Z=e^{iS_{{\cal N}=4}}\cdot W_{(k,0, \ldots ,0)},}
where $S_{{\cal N}=4}$ is the action of ${\cal N}=4$ SYM.
The expectation value of this operator can be computed by evaluating the classical action of the $D3_k$-brane.

\newsec{$D$-brane description of an Arbitrary Wilson loop}

In the previous section we have shown that Wilson loops labeled by Young tableaus with a single column are described by a $D5$-brane while  a $D3$-brane gives rise to tableaus with a single row. 
What is the gravitational description of Wilson loops in an arbitrary representation?

We now show that  given a Wilson loop operator described by an arbitrary Young tableau, that it can be described either in terms of a collection of giants or alternatively in terms of a collection of dual giants.

\subsec{Wilson loops as $D5$-branes}

In the previous section, we showed that the information about the number of  boxes in the 
Young tableau with one column is determined by the amount of fundamental string charge ending on the $D5$-brane.
For the case of a single $D5_k$-brane, this background electric charge is captured by inserting\insercharge\ 
\eqn\inserchargea{
\exp\left(-i k \int  dt\; \tilde{A}_0\right)}
in the  path integral of the defect conformal field theory. Equivalently, we can add 
the  Chern-Simons term:
\eqn\cherncharge{
-  \int dt\; k \tilde{A}_0.}
to the action \lag. This injects into the theory  a localized  external  particle of charge $k$ with respect to the $U(1)$ gauge field  $\tilde{A}_0$ on the $D5$-brane.

We  now show that describing half-BPS Wilson loop operators \Wilsonsusy\ labeled by tableaus
with more than one column corresponds to considering the brane configuration in \braneconf\ with 
multiple $D5$-branes.
 
 In order to show this,  we must consider the low energy effective field theory on $M$ $D5$-branes probing $N$ $D3$-branes. In this case, the $U(1)$ symmetry associated with  the $D5$-brane gets now promoted to a $U(M)$ symmetry, where $M$ is the number of $D5$-branes. 
Therefore, the defect conformal field theory  living on this brane configuration is given by\foot{For clarity, we write explicitly the indices associated with $U(N)$ and $U(M)$.}
\eqn\lagmult{
S=S_{{\cal N}=4}+\int dt\; i \chi_i^{I\dagger} \partial_t \chi_i^I+\chi_i^{I\dagger}(A_{0\hskip+1pt ij}+\phi_{ij})\chi_j^I,}
where $i,j$ is a  fundamental index of $U(N)$ while $I,J$ is a fundamental index of $U(M)$.

We need to understand how to realize in our defect conformal field theory that we have   $M$ $D5$-branes in AdS$_5\times$S$^5$ with a configuration of fundamental strings dissolved in them. Physically, the string endpoints introduce into  the system a background charge for the $U(M)$ gauge field which depends on the distribution of string charge among the $M$ $D5$-branes. The charge is  labeled by a representation $\rho=(k_1,\ldots,k_M)$ of $U(M)$, where now 
$\rho=(k_1,\ldots,k_M)$ is a Young tableau of $U(M)$. A charge in the representation $\rho=(k_1,\ldots,k_M)$ is produced when  $k_i$ fundamental strings end on the $i$-th $D5$-brane. This $D5$-brane configuration can be labeled by the array
 $(D5_{k_1},\ldots, D5_{k_M})$:
 \medskip
  \ifig\sphera{Array of strings producing a background charge  given by the representation $\rho=(k_1,\ldots,k_M)$ of $U(M)$. The $D5$-branes are drawn separated for illustration purposes only,  as they sit on top of each other.}{\epsfxsize1.8in\epsfbox{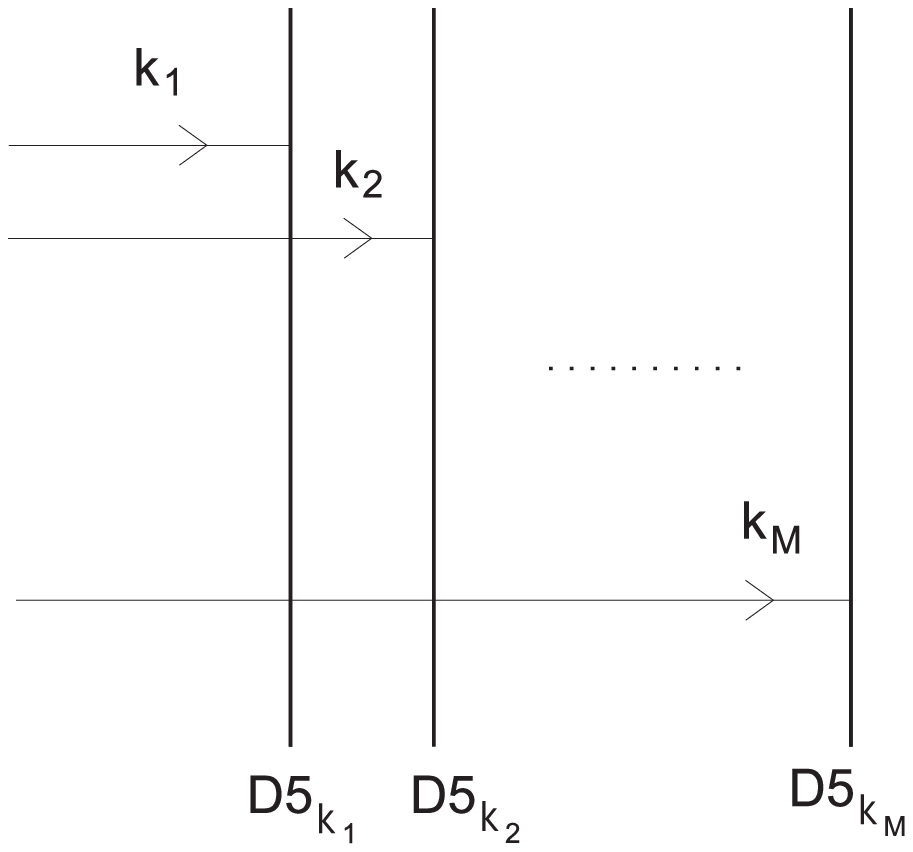}}

We must now add to the defect conformal field theory a term that captures that there is a static background charge $\rho=(k_1,\ldots,k_M)$ induced  in the system by the fundamental strings. This is accomplished by inserting into the path integral a Wilson loop operator for the gauge field ${\tilde A}_0$. The operator insertion is given by
\eqn\Wilsonsusya{
\hbox{Tr}_{(k_1,k_2,\ldots,k_M)}\;P \exp\left(-i\int dt\; \tilde{A}_0\right),}
which generalizes  \inserchargea\ when there are multiple $D5$-branes. We must also take into account the coupling of the localized fermions $\chi^I_i$ to $\tilde{A}_0$:
\eqn\extraextra{
S_{extra}=\int dt\; \chi_i^{I\dagger}\tilde{A}_{0\hskip+1pt IJ}\chi_i^J.}

In order to study what the $(D5_{k_1},\ldots, D5_{k_M})$ array in AdS$_5\times$S$^5$ corresponds to in ${\cal N}=4$ SYM, we need to calculate the following path integral
\eqn\pathnew{
Z=\int [D\chi ] [D\chi^\dagger] [D\tilde{A}_0] \; e^{i(S+S_{extra})}
\cdot \hbox{Tr}_{(k_1,k_2,\ldots,k_M)}\;P \exp\left(-i\int dt\; \tilde{A}_0\right),}
where $S$ is given in \lagmult\ and $S_{extra}$ in \extraextra.

We proceed by gauge fixing the $U(N)\times U(M)$ symmetry of the theory by diagonalizing $A_0+\phi$ and $\tilde{A}_0$ to have constant eigenvalues respectively. The eigenvalues are given by:
\eqn\eigenvals{\eqalign{
A_0+\phi&=\hbox{diag}(w_1,\ldots,w_N)\cr
\tilde{A}_0&=\hbox{diag}(\Omega_1,\ldots,\Omega_M).}}

Since the path integral in \pathnew\ involves integration over ${\tilde A}_0$ care must be taken in doing the gauge fixing procedure\foot{As discussed in footnote $11$, the gauge fixing associated with the $U(N)$ symmetry can be undone once one is done integrating out over $\chi$ and $\tilde{A}_0$.}. 
As shown in Appendix $C$, the measure over the Hermitean matrix ${\tilde A}_0$ combines 
with  the Fadeev-Popov determinant $\Delta_{FP}$ associated with  the gauge choice
\eqn\gaugechoice{
\tilde{A}_0=\hbox{diag}(\Omega_1,\ldots,\Omega_M)}
to yield the measure over a unitary matrix $U$. That is
\eqn\measurenew{
[D\tilde{A}_0]\cdot  \Delta_{FP}=[DU],}
with  $U=e^{i\beta {\tilde A}_0}$ and 
\eqn\measue{
[DU]=\prod_{I=1}^{M}{d\Omega_I}\;\Delta (\Omega)
\bar{\Delta}(\Omega),}
where $\Delta (\Omega)$ is the Vandermonde determinant\foot{There is a residual $U(1)^N$ gauge symmetry left over after the gauge fixing \gaugechoice\ which turns $\Omega_I$ into angular coordinates. We are then left with the proper integration domain over the angles of a unitary matrix.}:
\eqn\vander{
\Delta (\Omega)=\prod_{I<J}(e^{i\beta\Omega_I}-e^{i\beta\Omega_J}).}

In this gauge, another simplification occurs. The part of the action in \pathnew\ depending on the $\chi$ fields is given by:
\eqn\fermion{
\int dt\; \chi_i^{I\dagger}(\partial_t +w_i+\Omega_I)\chi_i^I.}
Correspondingly, the equations of motion are:
\eqn\motionnew{
(i\partial_t+w_i+\Omega_I)\chi_i^I=0\qquad \hbox{for}\qquad  i=1,\ldots,N\quad I=1,\ldots,M.}
Therefore, we have a system of $N\cdot M$ fermions  $\chi^I_i$ with energy $w_i+\Omega_I$. 

We can  explicitly integrate out the $\chi$ fields in $Z$ \pathnew\ by going to the Hamiltonian formulation, just as before. The fermion partition function is:
\eqn\partiman{\prod_{i=1}^N\prod_{J=1}^M(1+x_ie^{i\beta \Omega_J}),}
where as before $x_i=e^{i\beta w_i}$ is an eigenvalue of the holonomy matrix appearing in the Wilson loop operator \Wilsonsusy\  and $e^{i\beta \Omega_J}$ is an eigenvalue of the unitary matrix $U$.

Combining this with the computation of the measure, the path integral \pathnew\ can be written as 
\eqn\pathnewcalc{
Z= e^{iS_{{\cal N}=4}}\cdot \int [DU] \; \chi_{(k_1,\ldots,k_M)}(U^*)\prod_{i=1}^N\prod_{J=1}^M(1+x_ie^{i\beta \Omega_J}),}
where we have identified  the operator insertion \Wilsonsusya\ with a character in the $\rho=(k_1,\ldots,k_M)$ representation of $U(M)$:
\eqn\Wilsonsusyab{
  \chi_{(k_1,\ldots,k_M)}(U^*)\equiv \hbox{Tr}_{(k_1,\ldots,k_M)}e^{-i\beta\tilde{A}_0}.}

The partition function of the fermions \partiman\ can be expanded either in terms of characters of $U(N)$ or $U(M)$ by using  a generalization of  the formula we used in \expand. We find it convenient to write it in terms of characters of $U(M)$
\eqn\expandeor{
\prod_{J=1}^M(1+x_ie^{i\Omega_J})=\sum_{l=0}^M x_i^l\; \chi_{\matrix{(\underbrace{1, \ldots ,1},0,\ldots ,0)\cr \hskip-45pt l}}(U)=\sum_{l=0}^M x_i^lE_l(U_1,\ldots,U_M),}
where
\eqn\traces{
E_l(U)=\hbox{Tr}_{\matrix{(\underbrace{1, \ldots ,1},0,\ldots ,0)\cr \hskip-45pt l}}e^{i{\beta\tilde A}_0}}
is the character of $U(M)$ in the $l$-th antisymmetric product representation. We recall that $U=e^{i\beta \tilde{A}_0}$ and that $U_I=e^{i\beta\Omega_I}$ for $I=1,\ldots,M$ are its eigenvalues.

We now use the  following mathematical identity
\lref\BalantekinKM{
  A.~B.~Balantekin,
  ``Character Expansion For U(N) Groups And U(N/M) Supergroups,''
  J.\ Math.\ Phys.\  {\bf 25}, 2028 (1984).
}
\BalantekinKM 
\eqn\idenbalan{
\prod_{i=1}^N\sum_{l=0}^M x_i^l E_l(U)=\sum_{M\geq n_1\geq n_2\geq \ldots\geq n_N}\det(E_{n_j+i-j}(U))\;\chi_{(n_1,\ldots,n_N)}(x),}
where 
\eqn\further{
\chi_{(n_1,\ldots,n_N)}(x)=W_{(n_1,\ldots,n_N)}}
is precisely the half-BPS Wilson loop operator \Wilsonsusy\ in the $R=(n_1,\ldots,n_N)$ representation of $U(N)$. 
Therefore, the fermion partition function \partiman\ can be written in terms of $U(N)$ and $U(M)$ characters as follows
\eqn\idenbalan{
\prod_{i=1}^N\prod_{J=1}^M(1+x_ie^{i\beta \Omega_J})=\sum_{M\geq n_1\geq n_2\geq \ldots\geq n_N}\det(E_{n_j+i-j}(U))W_{(n_1,\ldots,n_N)}.}

The determinant $\det(E_{n_j+i-j}(U))$ can be explicitly evaluated by using Giambelli's formula (see e.g \groupthe )
\eqn\giam{
\det(E_{n_j+i-j}(U))=\chi_{(m_1,m_2,\ldots,m_M)}(U),}
where $\chi_{(m_1,m_2,\ldots,m_M)}(U)$ is the  character of $U(M)$ associated with the Young  tableau 
$(m_1,m_2,\ldots,m_M)$. This tableau is obtained from $(n_1,n_2,\ldots,n_N)$ by conjugation, which corresponds to transposing the tableau $(n_1,n_2,\ldots,n_N)$:

\ifig\sphera{A Young tableau and its conjugate. In the conjugate tableau the number of boxes in the $i$-th row is the number of boxes in the $i$-th column of the original one.}{\epsfxsize3.5in\epsfbox{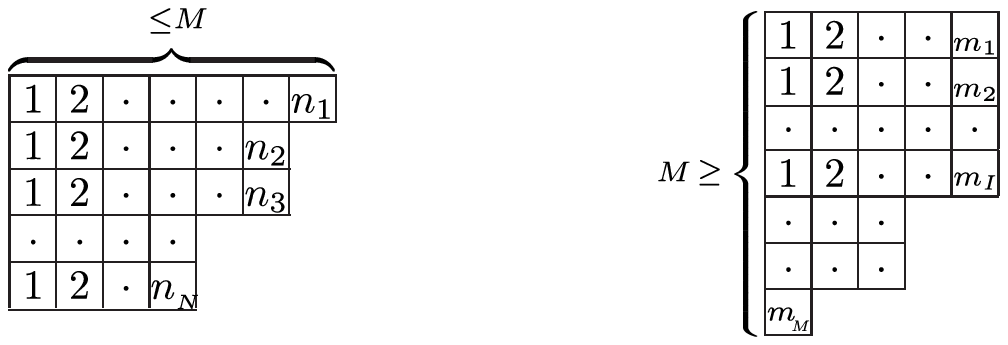}}







 The number of rows in the conjugated tableau $(m_1,m_2,\ldots,m_M)$ is constrained to be at most $M$ due to the  $M\geq n_1\geq n_2\geq \ldots\geq n_N$ constraint in the sum \idenbalan.

These computations allow us to write \pathnewcalc\ in the following way:
\eqn\pathfin{
Z=e^{iS_{{\cal N}=4}}\cdot \sum_{M\geq n_1\geq n_2\geq \ldots\geq n_N}W_{(n_1,\ldots,n_N)}\cdot
\int [DU]\; \chi_{(m_1,m_2,\ldots,m_M)}(U) \chi_{(k_1,\ldots,k_M)}(U^*).}
Now using orthogonality of $U(M)$ characters:
\eqn\ortho{
\int [DU]\; \chi_{(m_1,m_2,\ldots,m_M)}(U) \chi_{(k_1,\ldots,k_M)}(U^*)=\prod_{I=1}^M\delta_{m_I,k_I},}
we arrive at the final result
\eqn\pathfinfinfin{
Z=e^{iS_{{\cal N}=4}}\cdot W_{(l_1,\ldots,l_N)},}
where $(l_1,\ldots,l_N)$ is the tableau conjugate to $(k_1,\ldots,k_M)$. 

To summarize, we have shown that   a collection of $D5$-branes described by the array  $(D5_{k_1},\ldots, D5_{k_M})$ in AdS$_5\times$S$^5$ corresponds to the  half-BPS Wilson loop operator \Wilsonsusy\ in ${\cal N}=4$ SYM in the representation $R=(l_1,\ldots,l_N)$ of $U(N)$
\eqn\pathfinfinfin{
(D5_{k_1},\ldots, D5_{k_M})\longleftrightarrow Z=e^{iS_{{\cal N}=4}}\cdot W_{(l_1,\ldots,l_N)},}
 where $(l_1,\ldots,l_N)$ is the tableau conjugate to $(k_1,\ldots,k_M)$.
Thererefore, any half-BPS  Wilson loop operator  in ${\cal N}=4$ has a bulk realization.
We now move on to show that there is an alternative bulk formulation of  Wilson loop operators  in ${\cal N}=4$, now in terms of an array of $D3$-branes.

\subsec{Wilson loops as $D3$-branes}

Let's now consider the ${\cal N}=4$ gauge theory description of a configuration of multiple $D3$-branes 
in AdS$_5\times$S$^5$. As we have argued in section $4$, the only modification in the defect conformal field theory compared to the case with the $D5$-branes is to quantize the $\chi_i^I$ fields as chiral bosons as opposed to fermions. Therefore, we consider the defect conformal field theory action  \lagmult\ treating $\chi_i^I$  now as bosons.

Similarly to the case with multiple $D5$-branes, we realize the charge induced by the fundamental strings ending on the $D3$-branes by the Wilson loop operator \Wilsonsusya\ in the representation 
 $\rho=(k_1,\ldots,k_M)$ of $U(M)$, where  
$\rho=(k_1,\ldots,k_M)$ is a Young tableau of $U(M)$. A charge in the representation $\rho=(k_1,\ldots,k_M)$ is produced when  $k_i$ fundamental strings end on the $i$-th $D3$-brane. This $D3$-brane configuration can be labeled by the array
 $(D3_{k_1},\ldots, D3_{k_M})$:
 \medskip
  \ifig\sphera{Array of strings producing a background charge  given by the representation $\rho=(k_1,\ldots,k_M)$ of $U(M)$. The $D3$-branes are drawn separated for illustration purposes only,  as they sit on top of each other.}{\epsfxsize1.8in\epsfbox{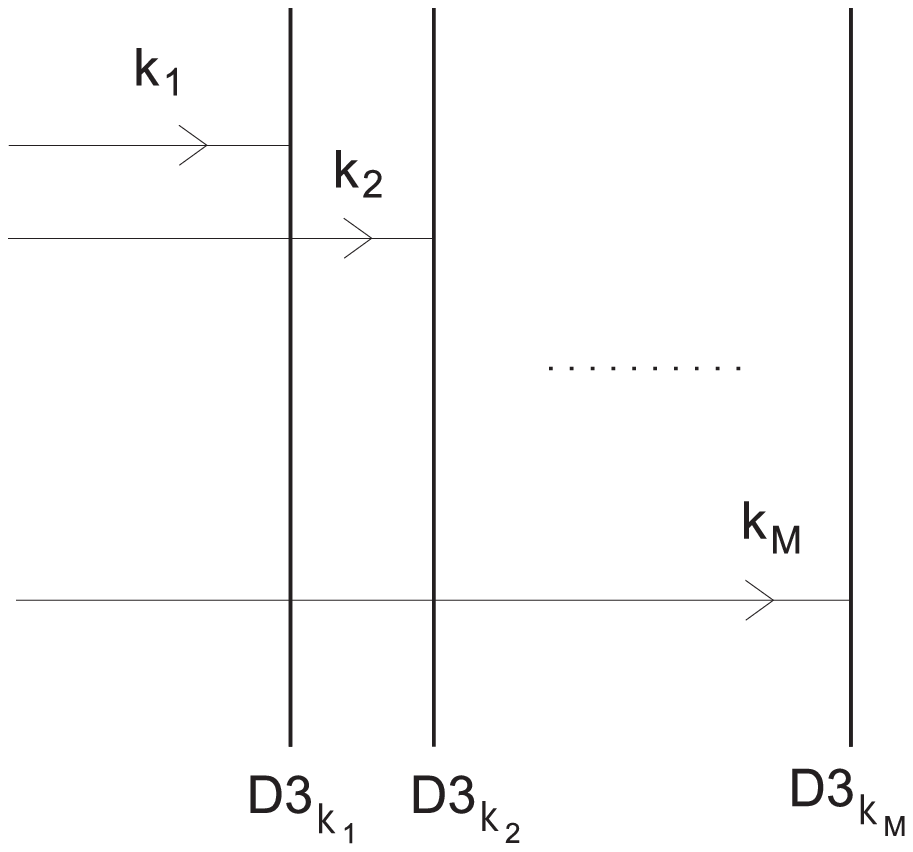}}  
\noindent
Therefore, in order to integrate out the degrees of freedom on the probe $D3$-branes we must calculate the path integral \pathnew\ treating $\chi_i^I$ as bosons.

We gauge fix the $U(N)\times U(M)$ as before. This gives us that $\chi_i^I$ are chiral bosons with energy $w_i+\Omega_I$. Their  partition function is then given by
\eqn\partybosons{
\prod_{i=1}^N\prod_{J=1}^M\left({1\over1-x_ie^{i\beta\Omega_J}}\right),}
where as before $x_i=e^{i\beta w_i}$ is an eigenvalue of the
holonomy matrix appearing in the Wilson loop operator.

Taking into account  the measure  change computed earlier, we have that 
\eqn\pathnewbose{ Z= e^{iS_{{\cal N}=4}}\cdot \int
[DU] \;
\chi_{(k_1,\ldots,k_M)}(U^*)\prod_{i=1}^N\prod_{J=1}^M\left({1\over1-x_ie^{i\beta\Omega_J}}\right),}where we have identified  the operator insertion \Wilsonsusya\ with a character in the $\rho=(k_1,\ldots,k_M)$ representation of $U(M)$:
\eqn\Wilsonsusyab{
  \chi_{(k_1,\ldots,k_M)}(U^*)\equiv \hbox{Tr}_{(k_1,\ldots,k_M)}e^{-i\beta\tilde{A}_0}.}

Now we use that the partition function of the bosons can be expanded
in terms of characters of $U(M)$ by generalizing formula \expandbose\
 \eqn\expandbosenew{
\prod_{J=1}^M\left({1\over1-x_ie^{i\beta\Omega_J}}\right)=\sum_{l=0}^\infty
x_i^l\; \chi_{\matrix{(l,0\ldots ,0)\cr \hskip-45pt
}}(U)=\sum_{l=0}^\infty x_i^lH_l(U_1,\ldots,U_M),}
 where
\eqn\tracebose{ H_l(U)=\hbox{Tr}_{\matrix{(l,0\ldots
,0)\cr \hskip-45pt }}e^{i\beta{\tilde A}_0}} 
is the character of $U(M)$ in the $l$-th symmetric product representation.

Using an identity from \BalantekinKM
 \eqn\identitycharbose{
\prod_{i=1}^N\sum_{l=0}^\infty x_i^lH_l(U)=\sum_{ n_1\geq n_2\geq
\ldots\geq n_N}\det(H_{n_j+i-j}(U))\ \chi_{(n_1,\ldots,n_N)}(x),}
where 
\eqn\wilbose{\chi_{(n_1,\ldots,n_N)}(x)=W_{(n_1,\ldots,n_N)}}
is the half-BPS Wilson loop operator corresponding to the Young tableau
$R=(n_1,\ldots,n_N)$ of $U(N)$.

The Jacobi-Trudy identity (see e.g \groupthe ) implies that 
 \eqn\giam{
\det(H_{n_j+i-j}(U))=\chi_{(n_1,n_2,\ldots,n_N)}(U),} where
$\chi_{(n_1,n_2,\ldots,n_N)}(U)$ is the character of $U(M)$ associated
with the Young  tableau $(n_1,n_2,\ldots,n_N)$. Considering the antisymmetry
of the elements in the same column, we get the constraint that $n_{M+1}=\ldots=n_{N}=0$, otherwise the
character vanishes.

These computations allow us to write  \pathnewbose\ as:
\eqn\pathfin{ Z=e^{iS_{{\cal N}=4}} \cdot \hskip-15pt\sum_{n_1\geq n_2\geq
\ldots\geq n_N}\hskip-10pt W_{(n_1,\ldots,n_N)}\cdot \int\;
[DU]\;\chi_{(n_1,\dots,n_N)}(U)\;{\chi}_{(k_1,\ldots,k_M)}(U^*).}
Using 
\eqn\ortbose{\int\;
[DU]\;\chi_{(n_1,\dots,n_N)}(U)\;{\chi}_{(k_1,\ldots,k_M)}(U^*)=\prod_{I=1}^{M}\delta_{n_I,k_I}\prod_{i=M+1}^N\delta_{n_i,0},} 
 we get that:
\eqn\pathfinfinbose{Z=e^{iS_{{\cal N}=4}} \cdot W_{(k_1,\ldots,k_M,\dots,0)}.}

We have shown that   a collection of $D3$-branes described by the array  $(D3_{k_1},\ldots, D3_{k_M})$ in AdS$_5\times$S$^5$ corresponds to the  half-BPS Wilson loop operator \Wilsonsusy\ in ${\cal N}=4$ SYM in the representation $R=(k_1,\ldots,k_N)$ of $U(N)$
\eqn\pathfinfinfin{
(D3_{k_1},\ldots, D3_{k_M})\longleftrightarrow Z=e^{iS_{{\cal N}=4}}\cdot W_{(k_1,\ldots,k_M,0,\dots,0)}.}
Therefore, any half-BPS  Wilson loop operator  in ${\cal N}=4$ has a bulk realization in terms of $D3$-branes.

To summarize, we have shown that a half-BPS Wilson loop described by an arbitrary Young tableau can be described in terms of a collection of $D5$-branes or $D3$-branes. We have shown that indeed  the relation between a Wilson loop in an arbitrary  representation  and a $D$-brane configuration  is precisely the one described in the introduction.

\bigbreak\bigskip\bigskip\centerline{{\bf Acknowledgements}}\nobreak
We would like to thank S. Ashok, Joaquim Gomis, A. Kapustin, L. Freidel, R. Myers, C. R\"omelsberger, G. Semenoff and  N. Suryanarayana for enjoyable discussions. F.P. would like to thank the Perimeter Institute for its hospitality. Research at the Perimeter Institute is supported in part by funds from NSERC of Canada and by MEDT of Ontario. 
The  work of F.P. is supported in part by the European Community's
Human Potential Programme under contract MRTN-CT-2004-005104
``Constituents, fundamental forces and symmetries of the universe",
by the FWO - Vlaanderen, project G.0235.05 and by the Federal Office
for Scientific, Technical and Cultural Affairs through the ``Interuniversity Attraction Poles Programme -- Belgian Science
Policy" P5/27.

 {\bf Note Added}. While this paper was getting ready for publication, the preprint 
\lref\YamaguchiTQ{
  S.~Yamaguchi,
  ``Wilson loops of anti-symmetric representation and D5-branes,''
  arXiv:hep-th/0603208.
}
\YamaguchiTQ\  appeared, which has overlap with parts of section $3$. In
\lref\HartnollHR{
  S.~A.~Hartnoll and S.~Prem Kumar,
  ``Multiply wound Polyakov loops at strong coupling,''
  arXiv:hep-th/0603190.
}
\HartnollHR\
an analogous $D5$-brane solution was considered for the AdS black hole background.

\appendix{A}{Supersymmetry of Wilson loops in ${\cal N}=4$ SYM}

In this Appendix  we study the constraints imposed by unbroken supersymmetry on the Wilson loop operators \Wilson\ of ${\cal N}=4$ SYM.  Previous studies of supersymmetry of  Wilson loops in ${\cal N}=4$ SYM include
\lref\DrukkerZQ{
  N.~Drukker, D.~J.~Gross and H.~Ooguri,
  ``Wilson loops and minimal surfaces,''
  Phys.\ Rev.\ D {\bf 60}, 125006 (1999)
  [arXiv:hep-th/9904191].
}
\lref\BianchiGZ{
  M.~Bianchi, M.~B.~Green and S.~Kovacs,
  ``Instanton corrections to circular Wilson loops in N = 4 supersymmetric
  JHEP {\bf 0204}, 040 (2002)
  [arXiv:hep-th/0202003].
}
\lref\ZaremboAN{
  K.~Zarembo,
  ``Supersymmetric Wilson loops,''
  Nucl.\ Phys.\ B {\bf 643}, 157 (2002)
  [arXiv:hep-th/0205160].
} \DrukkerZQ\BianchiGZ \ZaremboAN.

We want to  impose that the Wilson loop operator \Wilson\ is  invariant under
one-half of the ${\cal N}=4$ Poincare supersymmetries  and also
invariant under one-half of the conformal supersymmetries. The
Poincare supersymmetry transformations are given by
\eqn\susy{\eqalign{ \delta_{\epsilon_1}A_{\mu}&=i{\bar
\epsilon_1}\gamma_\mu\lambda\cr \delta_{\epsilon_1}\phi_{I}&=i{\bar
\epsilon_1}\gamma_I\lambda,}} while the superconformal supersymmetry
transformations are given
  \eqn\confsusy{\eqalign{
\delta_{\epsilon_2}A_{\mu}&=i{\bar
\epsilon_2}x^\nu\gamma_\nu\gamma_\mu\lambda\cr
\delta_{\epsilon_2}\phi_{I}&=i{\bar
\epsilon_2}x^\nu\gamma_\nu\gamma_I\lambda,}} where $\epsilon_{1,2}$
are ten dimensional Majorana-Weyl spinors of opposite chirality. The
use of ten dimensional spinors is  useful when comparing with  string
theory computations.

  Preservation of one-half of the Poincare supersymmetries locally at each point in the loop where the operator is defined yields:
 \eqn\susloop{
 P\epsilon_1=(\gamma_\mu \dot x^\mu+\gamma_I \dot y^I)\epsilon_1=0.}
Therefore, there are invariant spinors at each point in the loop if and
only if $\dot x^2+\dot y^2=0$. This requires that $x^\mu(s)$ is a
time-like curve and that $\dot y^I=n^I(s)\sqrt{-\dot x^2}$, where
$n^I(s)$ is a unit vector in $R^6$, satisfying  $n^2(s)=1$. Without loss of generality we
can perform a boost and put  the external particle labeling the loop at rest so that
the curve along $R^{1,3}$  is given by $(x^0(s),x^i(s)=0)$  and we
can also choose an affine parameter $s$ on the curve such that
$\sqrt{-\dot x^2}=1$.

In order for the Wilson loop to be supersymmetric, each point in the
loop must preserve the same spinor. Therefore, we must impose that
\eqn\glocal{ {dP(s)\over ds}=0,} which implies that $\ddot x^0=0$
and that $n^I(s)=n^I$. Therefore, supersymmetry selects a preferred
curve in superspace, the straight line Wilson loop operator, given
by \eqn\Wilsonsusyap{
 W_R(C)=\hbox{Tr}_{R}\;P \exp\left(i\int dt\; (A_0+\phi)\right),}
 where $\phi=n^I\phi_I$. The operators are now just labelled by a choice of   Young tableau $R$.
For future reference, we write explicitly  the 8 unbroken Poincare
supersymmetries.  They must satisfy \eqn\susya{ i{\bar
\epsilon_1}\gamma_0\lambda+in^I{\bar \epsilon_1}\gamma_I\lambda=0.}
Using relations for conjugation of spinor with the conventions used
here\eqn\rel{{\bar\chi}\zeta={\bar\zeta}\chi,\qquad\chi=\gamma^I\zeta\rightarrow{\bar\chi}=-{\bar\zeta}\gamma^I}
we arrive at

\eqn\unbrokensusy{ \gamma_0\gamma_I n^I\epsilon_1=\epsilon_1.}

In a similar manner it  is possible to prove that the straight line Wilson
loop operator \Wilsonsusyap\ also preserves one-half of the superconformal
supersymmetries.  The 8 unbroken superconformal supersymmetries are
given by:
 \eqn\unbrokensusycf{ \gamma_0\gamma_I
n^I\epsilon_2=-\epsilon_2.}

\appendix{B}{Supersymmetry of Fundamental String and of $D5_k$-brane}

In this Appendix we show that the particular embeddings  considered
for the fundamental string  and the $D5_k$-brane in section $3$  preserve half of the supersymmetries of the
background. We  will use  conventions similar to those in
\ref\SkenderisVF{
  K.~Skenderis and M.~Taylor,
  ``Branes in AdS and pp-wave spacetimes,''
  JHEP {\bf 0206}, 025 (2002)
  [arXiv:hep-th/0204054].
}.

For convinience we write again the metric we are interested in (we set $L=1$)
\eqn\ads{ ds^2_{AdS\times S}=u^2\eta_{\mu\nu}dx^\mu
dx^\nu+{du^2\over u^2} +d\theta^2+\sin^2\theta\;
d\Omega_4^2,} where the  metric on S$^4$ is given by: \eqn\mq{
d\Omega_4=d\varphi_1^2+\sin\varphi_1^2d\varphi_2^2+\sin\varphi_1^2\sin\varphi_2^2d\varphi_3^2+\sin\varphi_1^2\sin\varphi_2^2\sin\varphi_3^2d\varphi_4^2.}

It is useful to introduce tangent space gamma matrices, i.e.
$\gamma_{\underline{m}}=e^m_{\underline{m}}\Gamma_m$
$(m,\underline{m}=0,\ldots,9)$ where $e^m_{\underline{m}}$ is the
inverse vielbein and $\Gamma_m$ are the target space matrices:
 \eqn\flatm{\eqalign{\gamma_\mu={1\over
u}\Gamma_\mu\qquad(\mu=0,1,2,3),\qquad\gamma_4=u\Gamma_u,\qquad\gamma_5=\Gamma_\theta,\cr\gamma_{a+5}={1\over
\sin\theta}\left(\prod_{j=1}^{a-1}{1\over
\sin\varphi_j}\right)\Gamma_{\varphi_{a}}\qquad(a=1,2,3,4)}}

The Killing spinor of AdS$_5\times$S$^5$ in the coordinates \ads\ is given by \SkenderisVF\
 \eqn\ksp{\epsilon=\left[-u^{-{1\over
2}}\gamma_4h(\theta,\varphi_a)+u^{{1\over
2}}h(\theta,\varphi_a)(\eta_{\mu\nu}x^{\mu}\gamma^{\nu})\right]\eta_2+u^{{1\over
2}}h(\theta,\varphi_a)\eta_1}
 where
\eqn\h{h(\theta,\varphi_a)=e^{{1\over 2}\theta\gamma_{45}}e^{{1\over 2}\varphi_1\gamma_{56}}e^{{1\over 2}\varphi_2\gamma_{67}}e^{{1\over 2}\varphi_3\gamma_{78}}e^{{1\over 2}\varphi_4\gamma_{89}}}
$\eta_1$ and $\eta_2$ are constant ten dimensional  complex spinors with   negative
and positive ten dimensional  chirality, i.e.
\eqn\gael{\gamma_{11}\eta_1=-\eta_1\qquad\gamma_{11}\eta_2=\eta_2.}
They also satisfy: \eqn\gas{P_{-}\eta_1=\eta_1\qquad
P_{+}\eta_2=\eta_2} where $P_{\pm}={1\over 2}(1\pm i
\gamma^{0123})$. Thus, each spinor $\eta_{1,2}$ has  16 independent real
components. These can be written  in terms of ten dimensional Majorana-Weyl spinors $\epsilon_1$ and $\epsilon_2$ of negative and positive chirality respectively:
\eqn\etr{\eqalign{
\eta_1&=\epsilon_1-i \gamma^{0123}\epsilon_1\cr\eta_2&=\epsilon_2+i
\gamma^{0123}\epsilon_2.}}
By going to the boundary of AdS at $u\rightarrow \infty$, we can identify from \ksp\ $\epsilon_1$ as the Poincare supersymmetry parameter while $\epsilon_2$ is the superconformal supersymmetry parameter of ${\cal N}=4$ SYM.

The supersymmetries preserved by the embedding of
a probe, are those that satisfy
\eqn\unsusy{\Gamma_{\kappa}\epsilon=\epsilon} where
$\Gamma_{\kappa}$ is the  $\kappa$ symmetry  transformation matrix 
in the probe worldvolume theory  and $\epsilon$ is the Killing spinor of the
$AdS_5\times S_5$ background \ksp. Both $\Gamma_{\kappa}$ and $\epsilon$
have to be evaluated at the location of the probe.

Let's now consider  a fundamental string  with an AdS$_2$ worldvolume geometry with embedding:
 \eqn\embstr{
\sigma^0=x^0\qquad \sigma^1=u       \qquad x^i=0\qquad x^I=n^I.} The
position of the string on  the S$^5$ is parametrized by the five
constant angles
$(\theta,\varphi_1,\varphi_2,\varphi_3,\varphi_4)$ or alternatively by a unit vector $n^I$ in $R^6$.
The matrix $\Gamma_{\kappa}$ for a fundamental string with this embedding
reduces to 
\eqn\fo {\Gamma_{F1}=\gamma_{04}K} 
where $K$ acts on a
spinor $\psi$ by  $K\psi=\psi^{\ast}$. For later convenience we
define also the operator $I$ such that $I\psi=-i\psi$.

The equation \unsusy\  has to be satisfied at  every point on the string. Thus, the term proportional to $u^{{1\over 2}}$ gives:
\eqn\upm{\Gamma_{F1}h(\theta,\varphi_a)\eta_1=h(\theta,\varphi_a)\eta_1.}
The terms proportional to $u^{-{1\over 2}}$ and $u^{-{1\over 2}}x_0$
both give:
\eqn\upmb{{\Gamma}_{F1}h(\theta,\varphi_a)\eta_2=-h(\theta,\varphi_a)\eta_2.}
These can be rewritten as
\eqn\um{n^I\gamma_{0I}\eta_1=\eta_1^\ast\qquad
n^I\gamma_{0I}\eta_2=-\eta_2^\ast\qquad I=4,5,6,7,8,9} where
\eqn\vi{n^I(\theta,\varphi_1,\varphi_2,\varphi_3,\varphi_4)=\left(\eqalign{&\cos\theta\cr&\sin\theta\cos\varphi_1\cr&\sin\theta\sin\varphi_1\cos\varphi_2\cr&\sin\theta\sin\varphi_1\sin\varphi_2\cos\varphi_3\cr&\sin\theta\sin\varphi_1\sin\varphi_2\sin\varphi_3\cos\varphi_4\cr&\sin\theta\sin\varphi_1\sin\varphi_2\sin\varphi_3\sin\varphi_4}\right)=\left(\eqalign{&\cos\theta\cr&\sin\theta l^{\alpha}
 }\right),} \
where $\alpha=(5,6,7,8,9)$
and these vectors satisfy
$n^2=1$ and $l^2=1$. Considering the parametrization  \etr, the projection \um\  becomes:
\eqn\usf{\gamma_{0I}n^I\epsilon_1=\epsilon_1\qquad
\gamma_{0I}n^I\epsilon_2=-\epsilon_2.} We note
that $n^I$ define the position of the string in the $S_5$, so it
characterizes the unbroken rotational symmetry of the system. Therefore, the fundamental string preserves exactly the same supersymmetries as the Wilson loop operator \Wilsonsusyap.

We now study the $D5_k$-brane  embedding considered first by \PawelczykHY\CaminoAT:
\eqn\dfemb{ \sigma^0=x^0\qquad \sigma^1=u \qquad
\sigma^a=\varphi_a\qquad x^i=0\qquad
\theta=\theta_k=\hbox{constant}.}
There is an electric flux on the brane given by 
\eqn\elect{F_{04}=F=\cos\theta_k,}
where $k$ is the amount of fundamental string charge on the $D5_k$-brane. 

For this configuration, $\Gamma_{\kappa}$ is
\eqn\gammakdf{\eqalign{\Gamma_{D5}&={1\over \sqrt{1-F^2}}\gamma_{046789}KI+{F\over \sqrt{1-F^2}} \gamma_{6789}I\cr
&={1\over \sin\theta_k}\gamma_{046789}KI+{\cos\theta_k\over \sin\theta_k}\gamma_{6789}I}}

Following similar steps as for the fundamental string, we arrive at
\eqn\upm{\Gamma_{D5}h(\theta_k,\varphi_a)\epsilon_1=h(\theta_k,\varphi_a)\epsilon_1\qquad\bar{\Gamma}_{D5}h(\theta_k,\varphi_a)\epsilon_2=h(\theta_k,\varphi_a)\epsilon_2,}
where
\eqn\gammakdfb{\bar{\Gamma}_{D5}=-{1\over\sin\theta_k}\gamma_{046789}KI+{\cos\theta_k\over\sin\theta_k}\gamma_{6789}I.}

Using that $h^{-1}\gamma_{04}h=n^I\gamma_{0I}$ and that 
$h^{-1}\gamma_{6789}h=l^{\alpha}\gamma_{\alpha 56789}$ we have that the supersymmetry left unbroken by a $D5_k$-brane is given by:
 \eqn\usf{\gamma_{04}\epsilon_1=\epsilon_1\qquad
\gamma_{04}\epsilon_2=-\epsilon_2.} 
Therefore  it preserves the same supersymmetries  as  a fundamental string sitting at the north pole
(i.e $\theta=0$), labeled by the vector  $n^I=(1,0,0,0,0,0)$. This  vector selects the unbroken rotational
symmetry.

\appendix{C}{Gauge Fixing and the Unitary Matrix Measure}

In section  $5$ we have gauge fixed the $U(M)$ symmetry by imposing the diagonal, constant gauge:
\eqn\gaugeapp{
\tilde{A}_0=\hbox{diag}(\Omega_1,\ldots,\Omega_M).}
There is an associated Fadeev-Popov determinant $\Delta_{FP}$ corresponding to this gauge choice. This modifies the measure to
\eqn\measurenewapp{
[D\tilde{A}_0]\cdot \Delta_{FP},}
where now $[D\tilde{A}_0]$ involves integration only over the constant mode of the hermitean matrix $\tilde{A}_0$. 
Under an infinitessimal gauge transformation labelled by $\alpha$,  $\tilde{A}_0$ transforms by
\eqn\gaugetrans{
\delta\tilde{A}_0=\partial_{t}\alpha+i[\tilde{A}_0,\alpha],}
so that:
\eqn\fpp{
\Delta_{FP}=\hbox{det}\left(\partial_t\ +i[\tilde{A}_0,\ ]\right).}
An elementary computation yields
\eqn\fppres{
\Delta_{FP}=\prod_{l\neq 0}^\infty  {2\pi i l\over \beta}\prod_{I<J}\prod_{k=1}^{\infty}\left(1-{\beta^2(\Omega_I-\Omega_J)^2\over 4\pi^2k^2}\right),}
where we have introduced $\beta$ as an infrared regulator. Now, using the product representation of the $sin$ function we have  that up to an irrelevant constant:
\eqn\finfpp{
\Delta_{FP}=\prod_{I<J}4\; {\sin^2\left(\beta\left({\Omega_I-\Omega_J\over 2}\right)\right)\over (\Omega_I-\Omega_J)^2}.} 
This together with the formula for the measure of the  Hermitean matrix  $\tilde{A}_0$
\eqn\meashermite{
[D\tilde{A}_0]=\prod_{I<J}d\Omega_I (\Omega_I-\Omega_J)^2}
proves our claim that the gauge fixing effectively replaces the measure over the  Hermitean matrix $\tilde{A}_0$ by the measure over the unitary  $U=e^{i\beta \tilde{A}_0}$
\eqn\finunit{
[D\tilde{A}_0]\cdot \Delta_{FP}=[DU]= \prod_{I<J}d\Omega_I \Delta(\Omega)\bar{\Delta}(\Omega),}
where
\eqn\vanderfin{
\Delta(\Omega)=\prod_{I<J} (e^{i\beta\Omega_I}-e^{i\beta\Omega_J}).}

\listrefs

 \end